\documentclass[twocolumn]{autart}    % Enable this line and disable the 
\usepackage{graphicx}         
\usepackage{cite}
\usepackage{amsmath,amssymb,amsfonts}
\usepackage{algorithm}
\usepackage{algpseudocode}
\usepackage{xcolor}
\usepackage{bm}
\usepackage{longtable,tabularx}
\usepackage{multirow}
\usepackage{nomencl}  % for notation
\usepackage{cancel}
\usepackage{booktabs}

\DeclareMathAlphabet{\mathpzc}{OT1}{pzc}{m}{it}
\newcommand{\diag}{\mathrm{diag}}
\newcommand{\vect}{\mathrm{vec}}

\def\bf{\textbf}
\def\be{\begin{equation}}
\def\ee{\end{equation}}
\def\ba{\begin{array}}
\def\ea{\end{array}}

\errorcontextlines\maxdimen
\makeatletter
\newcommand*{\algrule}[1][\algorithmicindent]{\makebox[#1][l]{\hspace*{.5em}\vrule height .75\baselineskip depth .25\baselineskip}}%

\newcount\ALG@printindent@tempcnta
\def\ALG@printindent{%
    \ifnum \theALG@nested>0
        \ifx\ALG@text\ALG@x@notext
            \addvspace{-3pt}
        \else
            \unskip
            
            \ALG@printindent@tempcnta=1
            \loop
                \algrule[\csname ALG@ind@\the\ALG@printindent@tempcnta\endcsname]
                \advance \ALG@printindent@tempcnta 1
            \ifnum \ALG@printindent@tempcnta<\numexpr\theALG@nested+1\relax 
            \repeat
        \fi
    \fi
    }
\usepackage{etoolbox}
\patchcmd{\ALG@doentity}{\noindent\hskip\ALG@tlm}{\ALG@printindent}{}{\errmessage{failed to patch}}
\makeatother

\newcommand{\TR}[1]{{#1}}

\newcommand{\YL}[1]{{#1}}
 \newcommand{\YLnew}[1]{{#1}}

\begin{document}

\begin{frontmatter}

% \title{Dual Gaussian Process Based Model Predictive Control for quadrotors with Memory} 
% \title{\YL{Memory Based Model Predictive Control with Dual Gaussian Process}}
\title{Learning For Predictive Control: \\
A Dual Gaussian Process Approach}

\thanks[footnoteinfo]{This research was supported by the European Union within the framework of the National Laboratory for Autonomous Systems (RRF-2.3.1-21-2022-00002).}

\author[YH]{Yuhan Liu}\ead{y.liu11@tue.nl}, 
\author[PY]{Pengyu Wang}\ead{wangpy@kaist.ac.kr}, 
\author[YH,TR]{Roland T\'oth}\ead{r.toth@tue.nl}             

\address[YH]{Control Systems Group, Department of Electrical Engineering, Eindhoven University of Technology, %P.O. Box 513, 5600 MB 
Eindhoven, The Netherlands}                                         
\address[PY]{Flight Dynamics and Control Laboratory, Department of Aerospace Engineering, Korea Advanced Institute of Science and Technology, %291 Daehak-ro, 
Daejeon% 34141 
, Korea}  
\address[TR]{Systems and Control Laboratory, Institute for Computer Science and Control, %Kende u. 13-17, H-1111 
Budapest, Hungary}


\begin{abstract}                         % Abstract of not more than 200 words.
\YL{An important issue in {model-based} control {design} is that an accurate dynamic model of the system is generally nonlinear, complex, and costly to obtain. This limits achievable control performance in practice. Gaussian process (GP) based estimation {of system models} is an effective tool to learn unknown dynamics {directly} from input/output data. However, conventional GP-based control methods often ignore the computational cost associated with accumulating data during the operation of the system and how to handle forgetting in continuous adaption. In this paper, we present a novel Dual Gaussian Process (DGP) based model predictive control (MPC) strategy that enables {efficient use of online learning based predictive control}  %to have the advantages of both forgetting prevention and 
 {without the danger of catastrophic forgetting}. The bio-inspired DGP structure is a combination of a long-term GP and a short-term GP, where the long-term GP is used to keep the learned knowledge in memory and the short-term GP is employed to rapidly compensate unknown dynamics during online operation. Furthermore, a novel recursive online update strategy for the short-term GP is proposed to successively improve the learnt model during online operation. %in an efficient way. 
 {Effectiveness of the proposed strategy is demonstrated via numerical simulations.}}
\end{abstract}

\end{frontmatter}

\section{\YL{Introduction}}
{\emph{Model predictive control} (MPC)} has attracted considerable attention in recent decades due to its capability of dealing with control problems {under operational} constraints. {MPC} has {been applied in} a wide range of applications, including process control\cite{qin2003survey}, {flight control of quadrotors} \cite{aswani2013provably,elokda2021data}, and {motion control of} robots \cite{ding2019real}, etc. The performance of model predictive control is highly dependent on the accurate description of the system dynamics. Generally, the system dynamics can be modeled by first principle \TR{laws} such as {the} Newton-Euler approach. However, system{s} in real applications are usually {affected by} nonlinear couplings, time-varying disturbances, and unmodeled dynamics, {that may have a significant impact on the achievable control performance}. For example, aerodynamic effects such as complex interactions of rotor airflows with the ground and objects such as walls, air friction, and flapping dynamics of the rotor blades are \TR{hard to model and specific for each individual quadrotor}, but they {can seriously affect high-speed maneuvering}. \TR{As a result,} a comprehensive and precise model {for high performance operations} \TR{is often} costly to obtain.

Recently, various data-driven learning-based control methods have shown potential to {capture} unmodeled dynamics and achiev{e} superior performance in compensatory control over physical model-based control. {Collected} input-output data {from the system} can be {used} to {directly} construct an {accurate} data-driven model {by these methods}. %instead of an accurate model which is time-consuming to obtain through the first principle or identification.
Especially, \emph{Gaussian Process}  regression\cite{rasmussen2003gaussian}, which is a Bayesian nonparametric data-driven modeling method, is promising to automatically extract important features of the unknown dynamics from {measurement} data. %Gaussian process regression is a powerful nonparametric framework using  distribution over functions for nonlinear regression. 
Compar{ed to} other model learning methods {based on} \emph{Artificial Neural Networks} (ANNs)\cite{ren2009neural} and \emph{Support Vector Machines} (SVMs)\cite{laurain2015instrumental}, the main advantage of GP regression is that it does not only provide a single function estimate, but also a variance, i.e. confidence interval, which indicates the accuracy of the {model estimate}. Substantial results have been {achieved in} GP-based learning control. In \cite{beckers2019stable}, a GP-based data-driven dynamic model was identified for reducing the uncertainty of a given model and adapting the feedback gain by taking into account the model confidence. Furthermore, feedback linearization-based control had been also achieved by learning a control affine GP model \cite{umlauft2017feedback}. In \cite{helwa2019provably}, the GP was employed to estimate the error between the desired and actual accelerations, then a robust controller with uncertainty upper bound was proposed to guarantee the closed-loop stability. A $\mathcal{L}_1$ adaptive control with GP was presented in \cite{gahlawat2020l1}, where the unknown model was learnt by a GP to achieve the safe control objective without sacrificing  robustness. 

{Powerful results have also been achieved for} GP-MPC, {where} the unknown part of the system dynamics {is captured by a} learnt probabilistic GP model \YLnew{\cite{koller2018learning,hewing2019cautious,bonzanini2021learning,mesbah2022fusion}}. {While this provides flexibility to represent the unknown dynamics with their uncertainty bounds it also} introduces {a} stochastic term into the state propagation, {which in turn necessitates the use of approximations to make the control problem tractable} \YLnew{\cite{hewing2020learning}}. Another challenge is the high computational load during the receding horizon optimization, especially when the GP is employed to predict the function value based on a large data set. In general, GP is not suitable for large training sets because the computational cost to train a GP is cubic in terms of the number of data points. To solve this issue, various scalable GP methods have been developed to reduce the computational load, but preserve the prediction accuracy. The simplest way is to choose a \emph{subset of data} (SoD) of size $M\ll N$ randomly to represent the full training set. However, this strategy can lead to serious underfitting as the probability of discarding crucial information included in the full training set grows with the reduction rate. In \cite{lawrence2003fast}, instead of choosing \TR{a} subset randomly, an \emph{information vector machine} based method is proposed to select the subset more {efficiently}. Besides the SoD method, many {advanced} techniques  have been explored, including \TR{the} \emph{Nystr\"{o}m method}\cite{williams2001using}, \emph{deterministic training conditional approximation} (DTC)\cite{csato2002gaussian,seeger2003bayesian}, \emph{partially independent training conditional approximation} (PITC)\cite{tresp2000generalized}, and \emph{fully independent training conditional approximation} (FITC)\cite{snelson2005sparse}. A unifying framework of the mentioned algorithms was discussed in \cite{liu2020gaussian}.  
Several studies for GP-MPC were carried out with the applications of race cars\cite{hewing2019cautious,kabzan2019learning}, robot arms\cite{carron2019data}, and quadrotors\cite{cao2016gaussian}. 
However, most GP-based control methods do not perform online updating to avoid the issue of data bloating. {A}lthough{,} this would allow online adaptation with GP-based control during the system operation. By using data approximation techniques to counteract data bloating in online updating, the GP can gradually lose its knowledge learned during the training phase, and morph into a purely online adaptive regressor. Thus, designing a "memory"-based GP structure with online learning ability {is highly attractive}. 

Motivated by these facts and inspired by the biological concept of "long/short term memory" of human beings, we propose a novel \emph{dual Gaussian process} (DGP) based model predictive control strategy in this paper, as an extension of our preliminary work in the topic and presented in \cite{liu2021learning}. The DGP structure consists of two parts{: a} short-term GP {which} has the capability of online updat{ing} and is used for time-varying uncertainty compensation, and a long-term GP {that} is employed to learn the time-independent uncertainties and store the collected experience. The main contributions of this paper are summarized \TR{below.} 

1) By the authors' knowledge, the proposed DGP structure is the first framework that emphasizes both "memory" and "learning" ability by a combination of long/short-term GPs. As we show, such a method is capable of preventing forgetting relevant process information without the need of periodic reinforcement, but at the same time ensures exploration and adaptation to uncertain or varying  aspects of the dynamics. A long-term GP is utilized for the former while a short-term GP is used for the latter objectives. 

2) A novel recursive online updating strategy for the short-term GP is presented to continuously learn the unknown time-varying uncertainties during control operation. The proposed recursive online sparse GP has the ability of {efficiently} updating both the posterior mean and variance simultaneously, which especially benefits the MPC settings.

3) We show that the proposed DGP structure can be employed {in an} MPC scheme, which enables the conventional MPC to have additional bio-inspired advantages of online learning and \TR{remembering}. Also, rigorous results of DGP prediction at an uncertain input are provided in detail to permit the uncertainty propagation in GP-MPC scheme.

The remainder of this paper is organized as follows. Section {\ref{sec:2}} formalizes the model predictive control problem and introduces the preliminaries {on GPs}. The proposed recursive online sparse GP {regression} together with the novel dual GP structure are detailed in Section {\ref{sec:3}}. Then, {a} learning based predictive control strategy {based on the proposed} dual GP structure is presented in Section {\ref{sec:4}}. {To demonstrate the capabilities of the proposed adaptive predictive control scheme,} simulation examples {of adaptive flight control of a} quadrotor are given in Section {\ref{sec:5}}. {F}inally, conclusions on the achieved results are drawn in Section {\ref{sec:6}}.

% \YL{\textbf{Notations}}

\section{Preliminaries and Problem Formulation} \label{sec:2}
\subsection{\TR{The Data-generating System}}

In this paper, we consider discrete-time nonlinear systems that can be represented in the form of 
\begin{equation} \label{1}
    \bm{x}(k+1) = \bm{f}(\bm{x}(k),\bm{u}(k))+\bm{g}(\bm{x}(k),\bm{u}(k)) + \bm{v}(k),
\end{equation}
where $k\in\mathbb{Z}$ is the discrete time,  $\bm{x}(k)\in\mathbb{R}^{n_\mathrm{x}}$, $\bm{u}(k)\in\mathbb{R}^{n_\mathrm{u}}$, and  $\bm{v}(k)\in\mathbb{R}^{n_\mathrm{x}}$ are the state, control inputs, and \YLnew{noisy signals}, respectively, while $\bm{f}:\mathbb{R}^{n_\mathrm{x}\times n_\mathrm{u} } \to \mathbb{R}^{n_\mathrm{x}}$ and $\bm{g}:\mathbb{R}^{n_\mathrm{x}\times n_\mathrm{u} } \to \mathbb{R}^{n_\mathrm{x}}$ are bounded deterministic vector functions. The function $\bm{f}$ constitutes the physically well interpretable and a priori know dynamics of the system, i.e., its \emph{nominal model}, whereas, $\bm{g}$ represents the unknown, i.e., \emph{unmodeled} dynamics \YLnew{together with the external disturbances} of the system. \YLnew{The noisy signal} $\bm{v}$ is assumed to be an \emph{independent and identically distributed} (i.i.d.) white noise process \YLnew{$\bm{v}\sim\mathcal{N}(\mathbf{0},\bm{\Sigma}_\mathrm{\epsilon})$ where $\bm{\Sigma}_\mathrm{\epsilon} = \mathrm{diag}({\sigma}_{\epsilon,1}^2,\dots,{\sigma}_{\epsilon,n_\mathrm{x}}^2)$}. We moreover assume that the full state vector is available for measurement.

%\YL{{We consider} dynamic system{s} in this paper {that can be represented in the form of} %given by a general discrete-time nonlinear equation:
%\be
%\label{1}
%\bm{x}(k+1)=\bm{f}(\bm{x}(k),\bm{u}(k))+\bm{d}(k)
%\ee
%where $k\in\mathbb{Z}$ is the discrete time satisfying $t=kT_\mathrm{s}$ with sampling time $T_\mathrm{s}$ \textgreater $~0$, $\bm{x}(k)\in\mathbb{R}^{n_x}$ %and $\bm{x}(k+1)\in\mathbb{R}^{n_x}$ 
%is {the state variable} %, respectively, 
%$\bm{u}(k)\in\mathbb{R}^{n_\TR{\mathrm{u}}}$ {denotes} the control inputs, $\bm{d}(k)\in\mathbb{R}^{n_x}$ {represents} the external disturbances and unmodeled dynamics, $\bm{f}(\cdot):\mathbb{R}^{n_x}\times\mathbb{R}^{n_\TR{\mathrm{u}}}\to\mathbb{R}^{n_x}$ is a bounded nonlinear function, corresponding to the first principle model of the dyanamics.}

\YL{{Let us consider $\bm{f}$ to be linear allowing to write} \eqref{1} as
% \\[2mm]
% {I would get rid off $\bm{B}_d$, just complicates the problem and I would keep the noise process}
% \\[2mm]
\be
\label{2}
\bm{x}(k+1)=\bm{A}\bm{x}(k)+\bm{B}\bm{u}(k)+\bm{g}(\bm{x}(k),\bm{u}(k)) + \TR{\bm{v}(k)}
\ee
where $\bm{A}$ and $\bm{B}$ are the known system matrices derived from the linearization of the idealistic dynamics, hence $\bm{g}(\bm{x}(k),\bm{u}(k)) $ \TR{captures} the approximation error of the applied linearization and discretization or effects that cannot be captured reliably by the idealistic model \eqref{1} (i.e., both nonlinear effects and external disturbances).} 

\subsection{\TR{The Predictive Control Problem}}

\YL{In this paper, we consider a constrained finite-horizon optimal control problem for \eqref{2}. Given the current time instant $k \in \TR{\mathbb{Z}}$, time horizon $H \in \TR{\mathbb{N}}$, the state \TR{and input constraints} $\mathcal{X}\subseteq \mathbb{R}^{n_\TR{\mathrm{x}}}$ and $\mathcal{U}\subseteq \mathbb{R}^{n_\TR{\mathrm{u}}}$, \TR{respectively}, the goal is to find the sequence of admissible control inputs $\bm{U}_k=\{\bm{u}_{0|k},...,\bm{u}_{i|k}\}$ that satisfy $\bm{x}_{i|k}\in\mathbb{R}^{n_\TR{\mathrm{x}}}$ and $\bm{u}_{i|k}\in\mathbb{R}^{n_\TR{\mathrm{u}}}$ for all $i\in \TR{\mathbb{I}_0^H=\{i\in\mathbb{Z} \mid 0\leq i \leq H \}}$, that is, solving the following receding horizon optimization problem:
\be
\label{3}
\begin{aligned}
\min_{\bm{U}_k}~ &J(\bm{x}(k),\bm{U}_k)=\left(l_\mathrm{f}(\bm{x}_{H|k})+\!\!\sum_{i=0}^{H-1}l(\bm{x}_{i|k},\bm{u}_{i|k})\right)\\
\rm{s.t.}~  &\bm{x}_{0|k}=\bm{x}(k), ~~ \bm{u}_{0|k}=\bm{u}(k),\\
&\bm{x}_{i+1|k}=\bm{A}\bm{x}_{i|k}+\bm{B}\bm{u}_{i|k}+\bm{g}(\bm{x}_{i|k},\bm{u}_{i|k}),\\
&\bm{x}_{i|k}\in \mathcal{X}, ~\bm{u}_{i|k}\in \mathcal{U}.
\end{aligned}
\ee
where $l(\cdot):\mathbb{R}^{n_\TR{\mathrm{x}}}\times\mathbb{R}^{n_\TR{\mathrm{u}}}\to\mathbb{R}$ and $l_\mathrm{f}(\cdot):\mathbb{R}^{n_\TR{\mathrm{x}}}\to\mathbb{R}$ are the stage and terminal cost functions, which are defined in a quadratic form with weight matrices $\bm{Q}$, $\bm{Q}_\mathrm{f}$ and $\bm{R}$, \YLnew{i.e., $l(\bm{x},\bm{u})=\bm{x}^{\top}\bm{Q}\bm{x}+\bm{u}^{\top}\bm{R}\bm{u}$.}  Furthermore, we consider polytopic constraints for the state and control input\TR{:}
\begin{subequations}
\begin{align}
\label{c-1}
\mathcal{X}& =\{\bm{x}(k)|\bm{C}_x\bm{x}(k)\leq\bm{c}_x\} \\[1mm]
\label{c-2}
\mathcal{U} &=\{\bm{u}(k)|\bm{C}_u\bm{u}(k)\leq\bm{c}_u\}
\end{align}
\end{subequations}
where $\bm{C}_x\in\mathbb{R}^{n_\mathrm{c}\times n_\TR{\mathrm{x}}}$, $\bm{c}_x\in\mathbb{R}^{n_\mathrm{c}}$, ${n_\mathrm{c}}$ is the number of half-space constraints. \TR{The considered predictive control problem \eqref{3} can only be solved if $\bm{g}$ is known or if an approximate description of it is efficiently from data.}
}

\subsection{Learning with Gaussian Processes} \label{sec:sub:GP}
To guarantee the aforementioned state and control input constraints, \TR{a} model $\bm{\Delta}(\bm{x}(k),\bm{u}(k))$ \TR{of $\bm{g}$} should be identified reliably. Our goal is to construct a probabilistic model $\bm{\Delta}(\bm{x}(k),\bm{u}(k))$ from the system measurement data, and to improve accuracy gradually as more data are collected. We model  $\bm{\Delta}(\bm{x}(k),\bm{u}(k))$ as a GP where the state-input pairs $\bm{z}(k)\triangleq(\bm{x}^{\top}(k)\quad\bm{u}^{\top}(k))^{\top}\in\mathbb{R}^{n_\TR{\mathrm{x}}+n_\TR{\mathrm{u}}}$ and $\TR{\bm{y}(k)}\triangleq \bm{x}(k+1)-\bm{A}\bm{x}(k)-\bm{B}\bm{u}(k)$ are denoted as training inputs and outputs, respectively. \TR{In terms of definition, a vectorial \emph{Gaussian Process} $\mathcal{GP}: \mathbb{R}^{n_\mathrm{z}} \rightarrow \mathbb{R}^{n_\mathrm{x}}$ assigns to every point $\bm{z} \in \mathbb{R}^{n_\mathrm{z}}$ a random variable $\mathcal{GP}(\bm{z})$ taking values in $\mathbb{R}^{n_\mathrm{x}}$ such that, for any finite set $\{\bm{z}(\tau)\}_{\tau=1}^N \subset \mathbb{R}^{n_\mathrm{z}}$, the joint probability distribution of $\mathcal{GP}\left(\bm{z}(1)\right), \ldots, \mathcal{GP}\left(\bm{z}(N)\right)$ is multidimensional Gaussian. This constitutes a distribution over functions.} GPs are fully determined by
\be
\label{4}
\begin{aligned}
&\bm{\mu}(\bm{z})= \mathbb{E}[\bm{\Delta}(\bm{z})],\\
&\TR{\bm\kappa}(\bm{z},\bm{z}')= \mathbb{E}[(\bm{\Delta}(\bm{z})-\bm{\mu}(\bm{z}))(\bm{\Delta}(\bm{z}')-\bm{\mu}(\bm{z}'))^\TR{\top}],
\end{aligned}
\ee 
where $\bm{\mu}(\bm{z})$ is the mean function and $\YLnew{\bm{\kappa}}(\bm{z},\bm{z}')\triangleq \mathrm{cov}(\bm{\Delta}(\bm{z}),\bm{\Delta}(\bm{z}'))$ is the positive semi-definite covariance function which denotes a measure for the correlation (or similarity) of any two data points $(\bm{z}, \bm{z}')$. \YL{Then we assume that: 
\be
\label{5}
{\bm{\Delta}}(\bm{z})\sim\mathcal{GP}(\bm{\mu}(\bm{z}),\TR{\bm{\kappa}}(\bm{z},\bm{z}')),
\ee
\TR{which describes our prior belief over the uncertain dynamics.} The prior mean function is usually set to zero \TR{as all prior knowledge of the dynamics is assumed to be given in $f$}. \TR{Furthermore, GP regression is often implemented for each element of the GP-output variable separately, i.e, in terms of scalar-valued ${\Delta}_i\sim\mathcal{GP} ({\mu}_i, \kappa_i)$ with $i \in \mathbb{I}_1^{n_\mathrm{x}}$, constituting to $\bm{\Delta}=\vect(\Delta_1,\ldots,\Delta_{n_\mathrm{x}})$. While this approach greatly simplifies the estimation problem, it inherently results in the choice of $\bm{\kappa}=\diag(\kappa_1,\ldots,\kappa_{n_\mathrm{x}})$.} \TR{Regarding $\kappa_i$}, there is a wide selection of covariance functions, such as sinusoidal and Mat\'ern kernels, and in this paper we use the \emph{squared exponential function with automatic relevance determination} (SEARD):
\be
\label{6}
\TR{\kappa_i}\left(\bm{z}, \bm{z}'\right)=\YLnew{\sigma_{\mathrm{f},i}^{2} \exp \left(-\frac{1}{2}(\bm{z}- \bm{z}')^{\top}\bm{\Lambda}_{i}^{-1}(\bm{z}- \bm{z}')\right)}
\ee
as a prior due to its universal approximation capability, where $\YLnew{\sigma_{\mathrm{f},i}^{2}}\in\mathbb{R}_{\geq0}$ and \YLnew{$\bm{\Lambda}_i=\mathrm{diag}(\lambda_{i,1}^2,...,\lambda^2_\TR{i,n_\mathrm{z}})$} are signal variance and length-scale hyperparameters, respectively.} %\TR{For the sake of simplicity, we will take each $\kappa_i$ to be the same covariance functions, what we will denote by $\kappa$.}

Consider the training set $\mathcal{D}_\TR{N}= \TR{ \{(\bm{y}(k),\bm{z}(k)) \}_{k=1}^N}$ with $N$ noisy output $\bm{y}(k)=\bm{\Delta}(\bm{z}(k))+\YLnew{\bm{v}(k)}$. \TR{Let $\bm{Z}= \vect(\bm{z}(1),\ldots, \bm{z}(N))$ and $\bm{Y}_i=\YLnew{\vect({\bm{y}}_i(1),\ldots, \bm{y}_i(N))}$ for $i\in\mathbb{I}_{1}^{n_\mathrm{x}}$.} The Gaussian prior distribution for the function $\bm{\Delta}$ and the model likelihood  of the training set $\mathcal{D}_\TR{N}$ are given by:
\begin{subequations}
\label{7}
\begin{align}
P(\bm{\Delta}_\TR{i})&=\mathcal{N}(\bm{\Delta}_\TR{i}|\bm{0}, \bm{K}_{N,\TR{i}})  \label{7a}\\
P(\bm{Y}_\TR{i}|\TR{\bm{\Delta}_i}) &= \mathcal{N}(\bm{Y}_\TR{i}|\TR{\bm{\Delta}_i},\sigma_{\epsilon,\TR{i}}^2\bm{I}_{N}) \label{7b}
\end{align}
\end{subequations}
respectively, where $\bm{K}_{N,\TR{i}}\in\mathbb{R}^{N\times N}$ denotes the kernel matrix with \TR{elements} $\TR{[\bm{K}_\TR{N}]_{n,m}=\kappa_i(\bm{z}\TR{(n)},\bm{z}\TR{(m)})}$, \TR{$n,m\in\mathbb{I}_{1}^{N}$} and \TR{$\bm{\Delta}_i={\vect(\Delta_i(\bm{z}(1)),\ldots, \Delta_i(\bm{z}(N))}$ for $i\in\mathbb{I}_{1}^{n_\mathrm{x}}$}.
Thus, according to Bayes' rule, one can obtain the posterior distribution by \emph{maximum a posterior} (MAP) estimation
\begin{multline}
\label{8}
P(\bm{\Delta}|\bm{Y})\propto P(\bm{Y}|\bm{\Delta})P(\bm{\Delta}) \\
\propto \vect\bigl(\mathcal{N}(\bm{\Delta}_i|\bm{K}_{N,\TR{i}}(\bm{K}_{N,\TR{i}}+\sigma_{\epsilon\YLnew{,i}}^2\bm{I}_{N})^{-1}\bm{Y}_i, \\ (\bm{K}_{N,\TR{i}}^{-1}+\sigma_{\epsilon,\TR{i}}^{-2}\bm{I}_{N})^{-1}\bigr)
\end{multline}

\YL{Since the prior mean has been assumed \TR{to be} zero, one has the joint distribution over $\TR{\bm{\Delta}_i}$ and $\TR{\Delta_i^{*}}$:
\be
\label{9}
\left(\begin{array}{c}
\TR{\bm{\Delta}_i} \mid \bm{Z} \\
\TR{{\Delta}_i^{*}} \mid \bm{z}^*
\end{array}\right) \sim \mathcal{N}\left(\begin{array}{c}
\TR{\bm{\Delta}_i}  \\
\TR{\Delta_i^{*}}
\end{array} \bigg| \begin{array}{c}
\bm{0}
\end{array},\left[\begin{array}{cc}
\bm{K}_{N,\TR{i}} & \bm{K}_{N*,\TR{i}} \\
\bm{K}_{*N,\TR{i}} & ~k_{* *,\TR{i}}
\end{array}\right]\right)
\ee
where $[\bm{K}_{N*,\TR{i}}]_{j}={\kappa_\TR{i}}(\TR{\bm{z}(j)},{\bm{z}}^{*})$, $\bm{K}_{*N,\TR{i}}=\bm{K}_{N*,\TR{i}}^{\top}$ and ${k}_{**,\TR{i}}={\kappa_\TR{i}}({\bm{z}}^{*},{\bm{z}}^{*})$.}

Then, the associated posterior distribution of the function value $\bm{\Delta}({\bm{z}}^{*})$ at a new test point ${\bm{z}}^{*}$ can be derived by merging \eqref{8} and \eqref{9} together into:
\begin{align}
P(\bm{\Delta}^{*}|\mathcal{D}_\TR{N},\bm{z}^{*})&=\int P(\bm{\Delta}^{*}|\bm{Z},\bm{\Delta},\bm{z}^{*})P(\bm{\Delta}|\bm{Y})\mathrm{d}\bm{\Delta} \notag \\
% &=\TR{\vect}\bigl(\mathcal{N}(\mu_{\Delta,\TR{i}}(\bm{z}^{*}),\Sigma_{\Delta,\TR{i}}(\bm{z}^{*}))\bigr) \notag \\
&= \YLnew{\mathcal{N}(\bm{\mu}_{\Delta}(\bm{z}^{*}),\bm{\Sigma}_{\Delta}(\bm{z}^{*}))} \label{10}
\end{align}
where \YLnew{$\bm{\mu}_{\Delta}(\bm{z}^{*}) = \TR{\vect}(\mu_{\Delta,\TR{1}}(\bm{z}^{*}),...,\mu_{\Delta,\TR{n_\mathrm{z}}}(\bm{z}^{*}))$, $\bm{\Sigma}_{\Delta}(\bm{z}^{*})=\mathrm{diag}(\sigma^2_{\Delta,\TR{1}}(\bm{z}^{*}),...,\sigma^2_{\Delta,\TR{n_\mathrm{z}}}(\bm{z}^{*}))$}
with the predictive mean and variance \YLnew{for each dimension $i$}:
\begin{subequations}
\begin{align}
\label{11}
{\mu}_{\Delta,\TR{i}}(\bm{z}^{*})&=\bm{K}_{*N,\TR{i}}(\bm{K}_{N,\TR{i}}+\sigma_{\epsilon,i}^2\bm{I}_{N})^{-1}\bm{Y}_\TR{i} \\
%\ee
%\be
\label{12}
{\YLnew{\sigma^2_{\Delta,\TR{i}}}}(\bm{z}^{*})&={k}_{**,\TR{i}}\!-\!\bm{K}_{*N,\TR{i}}(\bm{K}_{N,\TR{i}}\!+\!\sigma_{\epsilon,i}^2\bm{I}_{N})^{-1}\bm{K}_{N*,\TR{i}}
\end{align}
\end{subequations}
\YL{Furthermore, given the marginal likelihood $P(\bm{Y}_\TR{i})= \mathcal{N}(\bm{0},\bm{K}_{N,\TR{i}}+\YLnew{\sigma_{\epsilon,i}}^2\bm{I}_{N})$ and the hyperparameters set $\YLnew{\bm{\theta}_i=\TR{\vect}(\sigma_{\epsilon,i}^{2},\sigma_{\mathrm{f},i}^{2},\lambda_{i,1},...,\lambda_{i,\TR{n_\mathrm{z}}})}$, the probabilistic GP model \eqref{5} can be trained by maximizing \TR{the likelihood of each individual predictive output distribution separately, i.e.,}  
\be
\label{13}
\begin{aligned}
\log P(\bm{Y}_\TR{i})= &-\frac{N}{2}\log 2\pi-\log|\YLnew{\bm{K}_{N,i}}+\YLnew{\sigma_{\epsilon,i}}^2\bm{I}_{N}|\\[1mm]
&-\frac{1}{2} \bm{Y}_\TR{i}^{\top}(\YLnew{\bm{K}_{N,i}}+\YLnew{\sigma_{\epsilon,i}}^2\bm{I}_{N})^{-1}\bm{Y}_\TR{i}
\end{aligned}
\ee
to find the optimal $\hat{\bm{\theta}_\TR{i}} \in \arg \max _{\bm{\theta}_\TR{i}} \log P(\bm{Y}_\TR{i})$ by means of conjugate gradient-based algorithm\cite{rasmussen2003gaussian}, which has a computational load of $\mathcal{O}(N^3)$ per iteration.}

\section{Recursive Online Sparse Gaussian {Regression} with Dual Structure} \label{sec:3}

\subsection{Sparse Variational Gaussian Process}

\TR{As we could see, due to the diagonal assumption of the prior covariance matrix $\bm{\kappa}$, the predictive distribution for each $\Delta_i$ can be independently formulated and trained. Hence, to avoid complexity of the notation, in this section we will drop the indexing for the output dimension of the GP as the derived results in this section can be used channel vise in the same manner as we have seen it in Section \ref{sec:sub:GP}.}

The computational load per test case for the full GP in~\eqref{11} and~\eqref{12} is $\mathcal{O}(N)$ and $\mathcal{O}(N^2)$ in terms of mean and variance, respectively. \YL{In this paper, we focus on the generalization of \emph{sparse variational GP regression} (SVGP), which has been first introduced in \cite{titsias2009variational} and further extended to mini-batch training \cite{hensman2013gaussian} and non-Gaussian likelihood \cite{hensman2015scalable} problems.}

The goal of sparse GP is to find a set of pseudo inputs $\bm{Z}_\mathrm{u}=\TR{\vect}(\bm{z}_{\mathrm{u},1},...,\bm{z}_{\mathrm{u},M})$ corresponding to pseudo outputs $\bm{\Delta}_\mathrm{u}=\TR{\vect}(\Delta_{\mathrm{u},1},...,\Delta_{\mathrm{u},M})$ of size $M\ll N$. %\TR{Let $\bm{\Delta}=\vect(\Delta)$}
\YL{In SVGP, the true posterior distribution $P(\bm{\Delta}|\mathcal{D}_\TR{N})$ is approximated by a Gaussian distribution $q(\bm{\Delta},\bm{\Delta}_\mathrm{u})=P(\bm{\Delta}|\bm{\Delta}_\mathrm{u})q(\bm{\Delta}_\mathrm{u})$ by means of variational inference, where a tractable variational distribution $q\left(\bm{\Delta}_\mathrm{u}\right)=\mathcal{N}(\bm{\Delta}_\mathrm{u}|\bm{m}_\mathrm{u},\bm{S}_\mathrm{u})$ is used,}
corresponding to the maximization of the \emph{evidence lower bound} (ELBO) of $\log P(\bm{Y})$:
\be
\label{14}
\begin{aligned}
\mathcal{L}(q) &\triangleq \int q\left(\bm{\Delta},\bm{\Delta}_\mathrm{u}\right) \log \frac{P\left(\bm{Y},\bm{\Delta},\bm{\Delta}_\mathrm{u}\right)}{q\left(\bm{\Delta},\bm{\Delta}_\mathrm{u}\right)} \mathrm{d} {\bm{\Delta}} \mathrm{d} \bm{\Delta}_\mathrm{u}\\
&=F(q)-\mathcal{KL}(q\left(\bm{\Delta}_\mathrm{u}\right)||P\left(\bm{\Delta}_\mathrm{u}\right)),
\end{aligned}
\ee
\YL{where $\mathcal{L}(q)$ is also known as the \emph{Variational Free Energy} (VFE) in variational learning and is used as an approximation of the log evidence for model selection\cite{friston2007variational}, }$F(q)=\int q\left(\bm{\Delta}_\mathrm{u}\right)P\left({\bm{\Delta}}|\bm{\Delta}_\mathrm{u}\right) \log P(\bm{Y} | {\bm{\Delta}}) \mathrm{d} {\bm{\Delta}}\mathrm{d} \bm{\Delta}_\mathrm{u}$, and $\mathcal{KL}(\cdot||\cdot)$ represents the Kullback-Leibler divergence which quantifies the similarity between two distributions. 
Then the optimal $q^{*}\left(\bm{\Delta}_\mathrm{u}\right)$ in terms of maximum of~\eqref{14} with mean $\bm{m}_\mathrm{u}$ and variance $\bm{S}_\mathrm{u}$ \TR{is}:
\begin{subequations} \label{15}
\begin{align}
\bm{m}_\mathrm{u}&=\sigma_{\epsilon}^{-2}\bm{S}_\mathrm{u} \bm{K}_{M}^{-1} \bm{K}_{M N}  \bm{Y}\\[1mm]
\bm{S}_\mathrm{u}&=\bm{K}_{M}\left(\bm{K}_{M}+\sigma_{\epsilon}^{-2} \bm{K}_{M N} \bm{K}_{N M}\right)^{-1} \bm{K}_{M}
\end{align} 
\end{subequations}
where $[\bm{K}_{MN}]_{i,j}=\TR{\kappa}(\bm{z}_{\mathrm{u},i},{\bm{z}\TR{(j)}})$ denotes the covariance matrix between pseudo inputs $\bm{Z}_\mathrm{u}$ and training inputs $\bm{Z}$, $\bm{K}_{MN} = \bm{K}_{MN}^{\top}$, $\bm{K}_{M}$ is the covariance matrix of \TR{the} pseudo inputs. See Appendix \ref{app:A} for details of the derivation of the optimal variational parameters. Merging the optimal distribution  $q^{*}\left(\bm{\Delta}_\mathrm{u}\right)$ into $\mathcal{L}(q)$, we get: 
\be
\label{16}
\mathcal{L}(q)\! =\! \log \mathcal{N}(\bm{Y}|\bm{0},\bm{Q}_{N}+\sigma_{\epsilon}^{2}\bm{I}_{N})\!-\!\frac{1}{2\sigma_{\epsilon}^{2}}\mathrm{tr}(\bm{K}_{N}\!-\!\bm{Q}_{N})
\ee 
where $\bm{Q}_{N}=\bm{K}_{NM}\bm{K}_{M}^{-1}\bm{K}_{MN}$. The pseudo inputs $\bm{Z}_\mathrm{u}$ and hyperparameters $\TR{\bm{\theta}}$ can be optimized by maximizing~\eqref{16}, which results in a computational load of $\mathcal{O}(NM^{2}+M^{3})$.

With the approximated posterior $q({\bm{\Delta}},\bm{\Delta}_\mathrm{u})$, the marginal distribution of ${\Delta}$ is:
\begin{multline}
\label{17}
q({\bm{\Delta}})=\int P\left({\bm{\Delta}}|\bm{\Delta}_\mathrm{u}\right) q\left(\bm{\Delta}_\mathrm{u}\right)\mathrm{d} \bm{\Delta}_\mathrm{u}\\
=\mathcal{N}({\mu}_{\Delta}(\bm{z}),\YLnew{{\sigma}^2_{\Delta}}(\bm{z})),
\end{multline}
\YL{Thus, the predictive mean and variance at the test point $\bm{z}^*$ \TR{are}
\begin{subequations}
\begin{align}
\label{18}
{\mu}_{\Delta}(\bm{z}^{*}) &=\bm{K}_{*M} \bm{K}_{M}^{-1}\bm{m}_\mathrm{u} \\[1mm]
\label{19}
\YLnew{{\sigma}^2_{\Delta}}(\bm{z}^{*}) &=k_{**}\!-\!\bm{K}_{*M}\left(\bm{K}_{M}^{-1}\!-\!\bm{K}_{M}^{-1}\bm{S}_\mathrm{u}\bm{K}_{M}^{-1}\right) \bm{K}_{M*} 
\end{align}
\end{subequations}
where $[\bm{K}_{*M}]_{j}=\TR{\kappa}(\bm{z}^{*},\bm{z}_{\mathrm{u},j})$.}

% weight space interpretation,
It is worth mentioning that, the hyperparameters and pseudo points are fixed after the ELBO maximization, and the GP is used to predict the mean and variance at a new data point during the operation. \YL{However, due to the fact that the external environment for real physical systems \TR{can be} complicated and time-varying, the offline pre-trained GP model is required to be successively improved to ensure continuous \TR{betterment/adaptation}.  }

\subsection{Recursive Online Sparse Gaussian Process }\label{sec:recGP}
Next a novel online update strategy for sparse Gaussian processes is proposed. The proposed method updates the posterior mean and variance of $q\left(\bm{\Delta}_\mathrm{u}\right)$ recursively based on the regression error at the current time step $k$.

On the basis of recursive Bayesian regression, and given a new online measurement output
\be
\label{m-1}
{y}\TR{(k)}={\Delta}({\bm{z}}\TR{(k)})+\YLnew{{v}\TR{(k)}},
\ee
the posterior mean and variance~\eqref{15} at the $k^\mathrm{th}$-step $q_k(\bm{\Delta}_\mathrm{u})=\mathcal{N}(\bm{\Delta}_\mathrm{u}|\bm{m}_\mathrm{u}^{k},\bm{S}_\mathrm{u}^{k})$ can be rewritten in terms of the posterior $q_{1:k-1}(\bm{\Delta}_\mathrm{u})=\mathcal{N}(\bm{\Delta}_\mathrm{u}|\bm{m}_\mathrm{u}^{k-1},\bm{S}_\mathrm{u}^{k-1})$ \TR{with}
\begin{subequations}
\begin{align}
\label{20}
&\bm{m}_\mathrm{u}^{k} = \bm{S}_\mathrm{u}^{k}(\bm{S}_\mathrm{u}^{k-1}\bm{m}_\mathrm{u}^{k}+\sigma_{\epsilon}^{-2}\bm{\Phi}_k^{\top}\TR{{y}(k)})\\[1mm]
&\bm{S}_\mathrm{u}^{k} = (\bm{S}_\mathrm{u}^{k-1}+\sigma_{\epsilon}^{-2}\bm{\Phi}_k^{\top}\bm{\Phi}_k)^{-1}
\end{align}
\end{subequations}
with kernel $\bm{\Phi}_k=\bm{K}_{zM} \bm{K}_{M}^{-1}|_{{\bm{z}}=\bm{z}\TR{(k)}}$ and $[\bm{\Phi}\TR{({\bm{z}}(k))}]_{j} = \phi(\TR{{\bm{z}}(k)},\TR{{\bm{z}}_\mathrm{u}(j)})$. Then~\eqref{18} can be seen as a linear combination of $M$ kernel functions, each one corresponding to a pseudo input:
{\setlength\abovedisplayskip{1pt}
 \setlength\belowdisplayskip{1pt}
\be
\label{21}
\YLnew{{\mu}}_{\Delta}(\bm{z})=\sum_{j=1}^{M}{m}_{\mathrm{u},j}\phi({\bm{z}},{\bm{z}}_{\mathrm{u},j})= \bm{\Phi}\TR{(\bm{z})} \bm{m}_\mathrm{u}
\ee
}
Note that~\eqref{21} can be interpreted as a weight-space representation of~\eqref{18}. Thus, recursive least squares \TR{can be efficiently used} to update~\eqref{18} and~\eqref{19} online \TR{under an incoming data} stream of \TR{measurement}  points. Given a new data point $\{{\bm{z}}\TR{(k)},  {y}\TR{(k)}\}$, the posterior mean and variance can be updated by:
\be
\label{22}
\left\{
\ba{cl}

\bm{m}_\mathrm{u}^{k} &=\bm{m}_\mathrm{u}^{k-1}+\bm{L}_k{r}_k\\

{r}_k &= {y}\TR{(k)}-\bm{\Phi}_k\bm{m}_\mathrm{u}^{k-1}\\

\bm{L}_{k} &= \bm{S}_\mathrm{u}^{k-1}\bm{\Phi}_k^{\top}\bm{G}_{k}^{-1}\\

\bm{G}_{k} &= \lambda + \bm{\Phi}_k\bm{S}_\mathrm{u}^{k-1}\bm{\Phi}_k^{\top}\\

\bm{S}_\mathrm{u}^{k} &= \lambda^{-1}(\bm{S}_\mathrm{u}^{k-1}-\bm{L}_{k}\bm{G}_{k}\bm{L}_{k}^{\top})
\ea\right.
\ee
\YL{where $0<\lambda\leq 1$ is the forgetting factor.}
The recursion starts from $\bm{m}_\mathrm{u}^{0}$ and $\bm{S}_\mathrm{u}^{0}$, which can be seen as the prior of the GP. This recursive online sparse GP, \TR{which is one of the main contributions of this paper,} can be embedded into the predictor of an MPC scheme, and is capable of dealing with varying disturbances and sudden changes in the  dynamics. 

% \floatname{algorithm}{\color{blue}Algorithm}
% \begin{algorithm}
% \caption{\YL{Recursive Sparse Online Gaussian Process Update Routine}}
% \label{alg:A}
% \begin{algorithmic}[1]
% \color{blue}
% \State \textbf{Input}:  Training data set $\mathcal{D}_0$, forgetting factor $\lambda$, simulation step size~$T_\mathrm{s}$, simulation time $T$,  ~$t=0$
% \State \textbf{Initialization}: Initial guess for $\bm{m}_{u}^{0}$ and $\bm{S}_{u}^{0}$ 

% \While{$t\leq T$ }  
     
%      \State Update the posterior of the short-term GP with the online measurement data according to \eqref{24};
%     %  \State Embed the predictive distribution \eqref{40} into the predictive model within the horizon $H$ ;
%     %  \State Solve the MPC problem \eqref{42} to obtain the optimal control input sequence;
%     %  \State Apply the first term of the optimal control input sequence on the system \eqref{2} and record data;

%  \EndWhile
% %  \State  Add the mission data into $\mathcal{D}_0$ and go back to Step 1. Wait for the next mission.

% \end{algorithmic}
% \end{algorithm}

\YL{\textbf{Remark 1}. (\textbf{Initial guess for $\bm{m}_\mathrm{u}^{0}$ and $\bm{S}_\mathrm{u}^{0}$.}) In \eqref{22}, the online update routine starts from $\bm{m}_\mathrm{u}^{0}$, which \TR{can be} provided by \TR{an} offline trained GP. The initial posterior variance $\bm{S}_\mathrm{u}^{0}$ is a user-defined parameter matrix and usually selected as $\bm{S}_\mathrm{u}^{0}=\beta^{-1}\bm{I}_\TR{M}$ with $0<\beta\leq 1$. According to the authors' experience, if the function space to be learnt is far away from the training data set $\mathcal{D}_\TR{N}$, one should choose a smaller $\beta$ indicating a smaller confidence level for the trained GP; Conversely, one chooses $\beta\approx1$.}

\YL{\textbf{Remark 2}. Comparing with the existing online GP methods, such as the (sparse online GP, SOGP) in \cite{csato2002sparse}, and the evolving GP in \cite{petelin2011control, kocijan2016modelling}, the proposed recursive online sparse Gaussian process is equivalent to directly updat\TR{ing} the posterior distribution $q(\bm{\Delta}_\mathrm{u})$ \TR{with} no need to re-optimize the hyperparameters or update the "dictionary" at every time step. Furthermore, both the posterior mean $\bm{m}_\mathrm{u}$ and variance $\bm{S}_\mathrm{u}$ will be updated simultaneously, which benefits the uncertainty propagation within the prediction horizon in \TR{an} MPC strategy.}

\subsection{Dual Gaussian Process Structure}\label{sec:dualGP}
\begin{figure}[t]
%   \captionsetup{font={small}}
  \includegraphics[width= 3.4in]{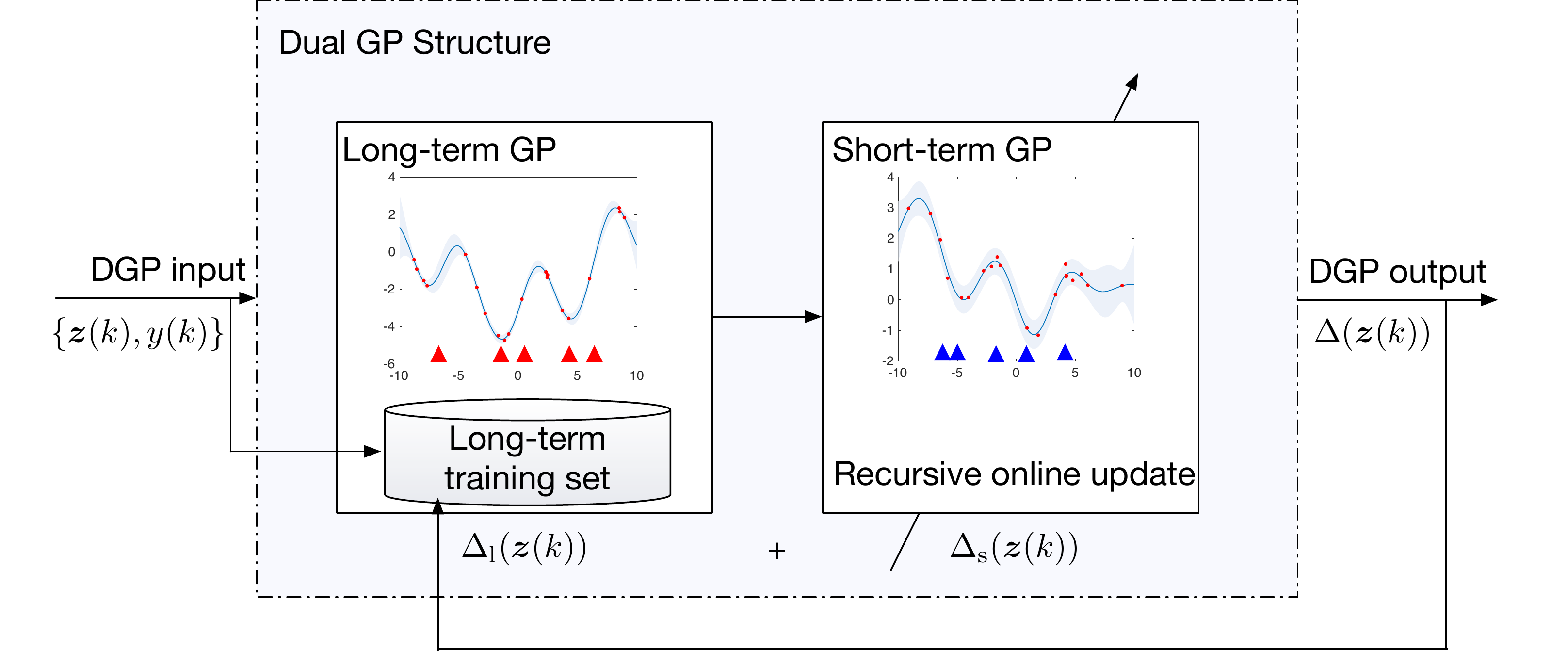}
  \caption{An illustration of the Dual Gaussian Process structure.} 
  \label{fig:2}
\end{figure}

Despite the adapting capability of the recursive online sparse GP, the learned dynamics obtained during the offline training phase, will be forgotten during the online learning phase. This means that, if the current evidence during the online control phase does not support the dynamics seen, they will gradually disappear via the forgetting factor $\lambda$  according to~\eqref{22}. For example, the "knowledge" for the uncertainties with large state bias will fade in case the state reaches equilibrium for a while.  Inspired by the biological concept of "long/short term memory", we present a novel structure named \emph{Dual Gaussian Process} (see Fig.~\ref{fig:2}), which allows to store collected experience and able to prevent knowledge forgetting. %The DGP structure is illustrated in Fig.~\ref{fig:2}.

The DGP structure mainly consists of two GPs: The first GP corresponds to the \emph{long-term} GP, which is employed to learn all the time-independent uncertainties and disturbances. The hyperparameters and the pseudo points of the long-term GP are kept fixed during the online learning phase, and will be batch updated after each mission. This makes the long-term GP \TR{to} have the ability to keep the \TR{accumulated experience} \TR{preserved in terms of a long-term memory, but still} evolve from mission to mission. On the other hand, the \emph{short-term} GP is utilized to adapt \TR{to} online time-varying disturbances and (sudden) changes in system dynamics. It is worth mentioning that, the short-term GP learns around the predicted mean value of the long-term GP, and the posterior is recursively updated during the online learning phase. Hence, \YLnew{a single dimension of} the unknown dynamics $\bm{\Delta}(\bm{z})$ can be rewritten as a sum of two GPs: 
\begin{subequations}
\begin{align}
\label{23}
%\begin{array}{c}
{\Delta}(\bm{z}) &= {\Delta}_\mathrm{l}(\bm{z})+{\Delta}_\mathrm{s}(\bm{z}) \\
{\Delta}_\mathrm{l}(\bm{z}) &\sim\mathcal{GP}({\mu}_\mathrm{l}(\bm{z}),{{\kappa}}(\bm{z},\bm{z}')),\\ \Delta_\mathrm{s}(\bm{z})&\sim\mathcal{GP}(0,\nu(\bm{z},\bm{z}'))
\end{align}
\end{subequations}
\noindent where the subscript \YLnew{$\mathrm{l}$} and \YLnew{$\mathrm{s}$} represent "long-term" and "short-term", respectively. {${{\nu}}(\bm{z},\bm{z}')$ denotes the covariance function for the short-term GP. For the simplicity of notation, the prior mean function is set as zero in the following derivation. Furthermore, the mean function ${\mu}_\mathrm{l}(\bm{z})$ for the long-term GP is obtained from the offline training phase}.

\YL{\textbf{Assumption 1}. The long-term GP ${\Delta}_\mathrm{l}$ and the short-term GP ${\Delta}_\mathrm{s}$ are independent with each other.} 

\YL{Note that Assumption 1 is mild because the sum of two functions, which are drawn from two arbitrary Gaussian processes, \TR{is also a} Gaussian process. Furthermore, the short-term GP is set to
learn around the mean value of the long-term GP, and the long-term GP \TR{is kept} fixed during the online update of the short-term GP.}

\YL{Then, considering the training set $\mathcal{D}_\TR{N}$, the prior distribution for ${\Delta}_\mathrm{l}$ and ${\Delta}_\mathrm{s}$ are given by:
\begin{subequations}\label{24}
\begin{align}
P(\bm{\Delta}_\mathrm{l})&=\mathcal{N}(\bm{\Delta}_\mathrm{l}\mid\bm{\mu}_\mathrm{l},\bm{K}_{N}),\\[1mm]
P(\bm{\Delta}_\mathrm{s})&=\mathcal{N}(\bm{\Delta}_\mathrm{s}\mid \bm{0},\bm{V}_{N}),
\end{align}
\end{subequations}
with \TR{$\bm{\Delta}_\mathrm{l}={\vect(\Delta_\mathrm{l}(\bm{z}(1)),\ldots, \Delta_\mathrm{l}(\bm{z}(N))}$ and $\bm{\Delta}_\mathrm{s}={\vect(\Delta_\mathrm{s}(\bm{z}(1)),\ldots, \Delta_\mathrm{s}(\bm{z}(N))}$} respectively. Moreover, the model likelihood is denoted as 
$P(\bm{Y}|\bm{\Delta}_\mathrm{l},\bm{\Delta}_\mathrm{s}) = \mathcal{N}(\bm{Y}|\bm{\Delta}_\mathrm{l}+\bm{\Delta}_\mathrm{s},\sigma_{\epsilon}^2\bm{I}_{N})$ where $\bm{V}_{N}$ represents the Gram matrix with $[\bm{V}]_{i,j}=\nu(\bm{z}\TR{(i)},\bm{z}\TR{(j)})$ for \TR{the} short-term GP.}

\begin{figure}[htpb]
% \captionsetup{font={small}}
  \centering\includegraphics[width= 3in]{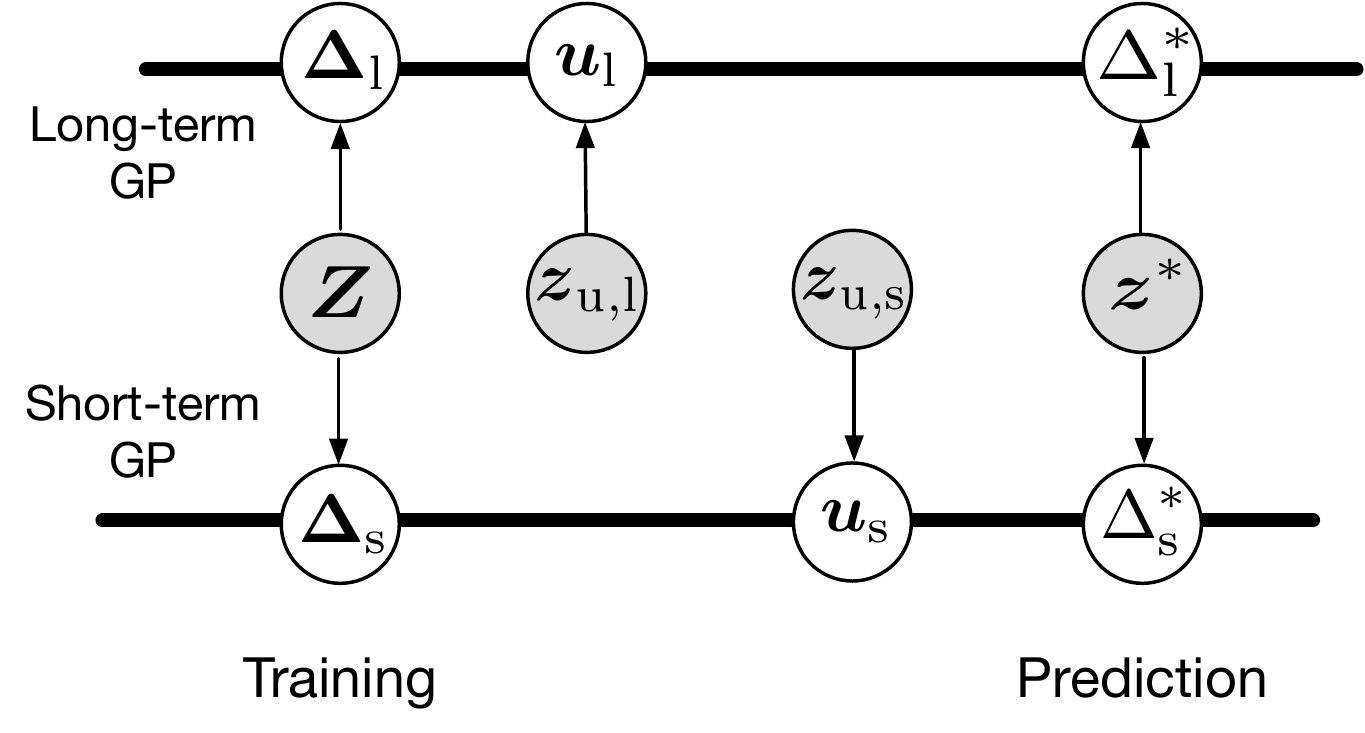}
% \centering\includegraphics[width= 0.5\textwidth]
\caption{Graphical model of \TR{the} DGP.}
\label{DGP_GM}
\end{figure}

To deal with large data sets, we introduce two sets of pseudo points for the DGP, i.e., $\bm{u}_{\mathrm{l},i}={\Delta}_\mathrm{l}(\bm{z}_{\mathrm{u},\mathrm{l},i})$ and $\bm{u}_{\mathrm{s},i}={\Delta}_\mathrm{s}(\bm{z}_{\mathrm{u},\mathrm{s},i})$ where $\bm{z}_{(\cdot),i}$ denote the \TR{$M$ number of} pseudo inputs for the long/short-term GP, respectively. The graphical model for DGP is shown in Fig. \ref{DGP_GM}.

\YL{Subsequently, one can obtain the following conditional distributions:
\begin{subequations}
\label{25}
\begin{align}
P(\bm{\Delta}_\mathrm{l}|\bm{u}_\mathrm{l}) &= \mathcal{N}(\bm{\Delta}_\mathrm{l}|\bm{K}_{NM}\bm{K}_{M}^{-1}\bm{u}_\mathrm{l}, \bm{K}_{N}- \bm{Q}_{N}) \\[1mm]
P(\bm{\Delta}_\mathrm{s}|\bm{u}_\mathrm{s}) &= \mathcal{N}(\bm{\Delta}_\mathrm{s}|\bm{V}_{NM}\bm{V}_{M}^{-1}\bm{u}_\mathrm{s}, \bm{V}_{N}- \bm{O}_{N})
\end{align}
\end{subequations}
where $\bm{O}_{N}=\bm{V}_{NM}\bm{V}_{M}^{-1}\bm{V}_{MN}$. Also, the marginal likelihood of DGP gives as:
\begin{multline}\label{26}
\log P(\bm{Y}) =\log \int P(\bm{Y}|\bm{\Delta}_\mathrm{l},\bm{\Delta}_\mathrm{s})P(\bm{\Delta}_\mathrm{l}|\bm{u}_\mathrm{l})P(\bm{\Delta}_\mathrm{s}|\bm{u}_\mathrm{s})\\
 P(\bm{u}_\mathrm{l})P(\bm{u}_\mathrm{s})\mathrm{d}\bm{\Delta}_\mathrm{l}\mathrm{d}\bm{\Delta}_\mathrm{s}\mathrm{d}\bm{u}_\mathrm{l}\mathrm{d}\bm{u}_\mathrm{s}
\end{multline}
}
Next, two variational distributions $q(\bm{u}_\mathrm{l})=\mathcal{N}(\bm{u}_\mathrm{l}|\bm{m}_{\mathrm{u},\mathrm{l}},\bm{S}_{\mathrm{u},\mathrm{l}})$ and $q(\bm{u}_\mathrm{s})=\mathcal{N}(\bm{u}_\mathrm{s}|\bm{m}_{\mathrm{u},\mathrm{s}},\bm{S}_{\mathrm{u},\mathrm{s}})$ are introduced to approximate the true  posterior: 
\begin{multline} \label{27}
P(\bm{\Delta}_\mathrm{l},\bm{\Delta}_\mathrm{s},\bm{u}_\mathrm{l},\bm{u}_\mathrm{s}|\bm{Y})\approx \\ P(\bm{\Delta}_\mathrm{l}|\bm{u}_\mathrm{l})P(\bm{\Delta}_\mathrm{s}|\bm{u}_\mathrm{s})q(\bm{u}_\mathrm{l})q(\bm{u}_\mathrm{s}).
\end{multline}
Then, the marginalized distributions of $\bm{\Delta}_\mathrm{l}$ and $\bm{\Delta}_\mathrm{s}$
\begin{subequations}
\label{28}
\begin{align}
q(\bm{\Delta}_\mathrm{l}) &= \int P(\bm{\Delta}_\mathrm{l}|\bm{u}_\mathrm{l})q(\bm{u}_\mathrm{l})\mathrm{d}\bm{u}_\mathrm{l}\\[1mm]
q(\bm{\Delta}_\mathrm{s}) &= \int P(\bm{\Delta}_\mathrm{s}|\bm{u}_\mathrm{s})q(\bm{u}_\mathrm{s})\mathrm{d}\bm{u}_\mathrm{s}
\end{align}
\end{subequations}
can be simply computed as %according to the Gaussian identities:
{\setlength\abovedisplayskip{3pt}
 \setlength\belowdisplayskip{3pt}
\begin{subequations}
\label{29}
\begin{align}
q(\bm{\Delta}_\mathrm{l}) &= \mathcal{N}(\overbrace{\bm{K}_{NM}\bm{K}_{M}^{-1}\bm{m}_{\mathrm{u},\mathrm{l}}}^{\bm{m}_{\mathrm{l}}},\overbrace{\bm{K}_{N}+\hat{\bm{Q}}_{N}}^{\bm{S}_{\mathrm{l}}})\\[0.5mm]
q(\bm{\Delta}_\mathrm{s}) &= \mathcal{N}(\underbrace{\bm{V}_{NM}\bm{V}_{M}^{-1}\bm{m}_{\mathrm{u},\mathrm{s}}}_{\bm{m}_{\mathrm{s}}},\underbrace{\bm{V}_{N}+\hat{\bm{O}}_{N}}_{\bm{S}_{\mathrm{s}}})
\end{align}
\end{subequations}
\noindent where $\hat{\bm{Q}}_{N}= \bm{K}_{NM}\bm{K}_{M}^{-1}(\bm{S}_{\mathrm{u},\mathrm{l}}-\bm{K}_{M})\bm{K}_{M}^{-1}\bm{K}_{MN}$, and  $\hat{\bm{O}}_{N}= \bm{V}_{NM}\bm{V}_{M}^{-1}(\bm{S}_{\mathrm{u},\mathrm{s}}-\bm{V}_{M})\bm{V}_{M}^{-1}\bm{V}_{MN}$. 

It is worth mentioning that the posterior for long-term GP $q(\bm{u}_\mathrm{l})$ is obtained after the offline training phase and it is fixed during the inference. Furthermore, the posterior $q(\bm{u}_\mathrm{s})$ for the short-term GP is updated with~\eqref{24}.
Then we can obtain the lower bound of $\log P(\bm{Y})$ for DGP as follows:
\begin{multline}
\label{30}
%\begin{aligned}
\mathcal{L}(q) \triangleq \int q\left(\bm{\Delta}_\mathrm{l},\bm{\Delta}_\mathrm{s},\bm{u}_\mathrm{l},\bm{u}_\mathrm{s}\right) \log \frac{P\left(\bm{Y}, \bm{\Delta}_\mathrm{l},\bm{\Delta}_\mathrm{s},\bm{u}_\mathrm{l},\bm{u}_\mathrm{s}\right)}{q\left(\bm{\Delta}_\mathrm{l},\bm{\Delta}_\mathrm{s},\bm{u}_\mathrm{l},\bm{u}_\mathrm{s}\right)}\\
\quad\quad\mathrm{d}\bm{\Delta}_\mathrm{l}\mathrm{d}\bm{\Delta}_\mathrm{s}\mathrm{d}\bm{u}_\mathrm{l}\mathrm{d}\bm{u}_\mathrm{s}\\
=\int q(\bm{u}_\mathrm{l})\int P(\bm{\Delta}_\mathrm{s}|\bm{u}_\mathrm{s})q(\bm{u}_\mathrm{s})\int P(\bm{\Delta}_\mathrm{l}|\bm{u}_\mathrm{l})\\
\quad \log P(\bm{Y}|\bm{\Delta}_\mathrm{l},\bm{\Delta}_\mathrm{s})\mathrm{d}\bm{\Delta}_\mathrm{l}\mathrm{d}\bm{\Delta}_\mathrm{s}\mathrm{d}\bm{u}_\mathrm{s}\mathrm{d}\bm{u}_\mathrm{l}\\
-\mathcal{KL}(q\left(\bm{u}_\mathrm{l}\right)||P\left(\bm{u}_\mathrm{l}\right))-\mathcal{KL}(q\left(\bm{u}_\mathrm{s}\right)||P\left(\bm{u}_\mathrm{s}\right))
%\end{aligned}
\end{multline}
Due to the fact that the posterior of the long-term GP is fixed during the online learning phase, we can equate the variational parameters of $q(\bm{u}_\mathrm{l})$ to~\eqref{15}:  $\bm{m}_{\mathrm{u},\mathrm{l}}=\bm{m}_\mathrm{u}$ and $\bm{S}_{\mathrm{u},\mathrm{l}}=\bm{S}_\mathrm{u}$. Next we will compute the analytic form of~\eqref{30} to get the optimal variational parameters for $q(\bm{u}_\mathrm{s})$. The inner integral for the first term of~\eqref{30} can be solved as:
{\setlength\abovedisplayskip{3pt}
\be 
\label{31}
\begin{aligned}
&\int P(\bm{\Delta}_\mathrm{l}|\bm{u}_\mathrm{l}) \log P(\bm{Y}|\bm{\Delta}_\mathrm{l},\bm{\Delta}_\mathrm{s})\mathrm{d}\bm{\Delta}_\mathrm{l} \\
&=\int\mathcal{N}(\bm{\Delta}_\mathrm{l}|\bm{K}_{NM}\bm{K}_{M}^{-1}\bm{u}_\mathrm{l}, \bm{K}_{N}-\bm{Q}_{N})
\log P(\bm{Y}|\bm{\Delta}_\mathrm{l},\bm{\Delta}_\mathrm{s})\mathrm{d}\bm{\Delta}_\mathrm{l}\\
&=\log \mathcal{N}(\bm{Y}|\bm{K}_{NM}\bm{K}_{M}^{-1}\bm{u}_\mathrm{l}+\bm{\Delta}_\mathrm{s},\sigma_{\epsilon}^2\bm{I}_{N})-\frac{1}{2\sigma_{\epsilon}^2}\mathrm{tr}(\bm{K}_{N}-\bm{Q}_{N})
\end{aligned}
\ee
}
By merging the inner integral into the second integral in~\eqref{32}:
\begin{multline}
\label{32}
%\begin{aligned}
\int P(\bm{\Delta}_\mathrm{s}|\bm{u}_\mathrm{s})q(\bm{u}_\mathrm{s})\int P(\bm{\Delta}_\mathrm{l}|\bm{u}_\mathrm{l})\log P(\bm{Y}|\bm{\Delta}_\mathrm{l},\bm{\Delta}_\mathrm{s})\mathrm{d}\bm{\Delta}_\mathrm{s}\mathrm{d}\bm{u}_\mathrm{s}\\
=\int q(\bm{\Delta}_\mathrm{s})\log \mathcal{N}(\bm{Y}|\bm{K}_{NM}\bm{K}_{M}^{-1}\bm{u}_\mathrm{l}+\bm{\Delta}_\mathrm{s},\sigma_{\epsilon}^2\bm{I}_{N})\mathrm{d}\bm{\Delta}_\mathrm{s}\\
\quad-\frac{1}{2\sigma_{\epsilon}^2}\mathrm{tr}(\bm{K}_{N}- \bm{Q}_{N})\\
=\log \mathcal{N}(\bm{Y}|\bm{K}_{NM}\bm{K}_{M}^{-1}\bm{u}_\mathrm{l}+\bm{m}_\mathrm{s},\sigma_{\epsilon}^2\bm{I}_{N})\\
\quad-\frac{1}{2\sigma_{\epsilon}^2}\mathrm{tr}(\bm{K}_{N}- \bm{Q}_{N})-\frac{1}{2\sigma_{\epsilon}^2}\mathrm{tr}(\bm{S}_\mathrm{s}).
%\end{aligned}
\end{multline}
}
% \vspace{-22pt}
Substituting~\eqref{32} back to~\eqref{30}, one has:
{\setlength\abovedisplayskip{3pt}
 \setlength\belowdisplayskip{3pt}
\begin{align}
\label{33}
\notag \mathcal{L}(q)&=\int q(\bm{u}_\mathrm{l})\log \mathcal{N}(\bm{Y}|\bm{K}_{NM}\bm{K}_{M}^{-1}\bm{u}_\mathrm{l}+\bm{m}_\mathrm{s},\sigma_{\epsilon}^2\bm{I}_{N})\mathrm{d}\bm{u}_\mathrm{l}\\ \notag
&\quad-\mathcal{KL}(q\left(\bm{u}_\mathrm{l}\right)||P\left(\bm{u}_\mathrm{l}\right))-\mathcal{KL}(q\left(\bm{u}_\mathrm{s}\right)||P\left(\bm{u}_\mathrm{s}\right))\\  
&\quad-\frac{1}{2\sigma_{\epsilon}^2}\mathrm{tr}(\bm{K}_{N}- \bm{Q}_{N})-\frac{1}{2\sigma_{\epsilon}^2}\mathrm{tr}(\bm{S}_\mathrm{s}) \\ \notag
&\leq \log\int\mathcal{N}(\bm{Y}|\bm{K}_{NM}\bm{K}_{M}^{-1}\bm{u}_\mathrm{l}+\bm{m}_\mathrm{s},\sigma_{\epsilon}^2\bm{I}_{N})P\left(\bm{u}_\mathrm{l}\right)\mathrm{d}\bm{u}_\mathrm{l}\\[-1mm]\notag
&-\mathcal{KL}(q\left(\bm{u}_\mathrm{s}\right)||P\left(\bm{u}_\mathrm{s}\right))-\frac{1}{2\sigma_{\epsilon}^2}\mathrm{tr}(\bm{K}_{N}-\bm{Q}_{N})-\frac{1}{2\sigma_{\epsilon}^2}\mathrm{tr}(\bm{S}_\mathrm{s})
\end{align}
}
Using Gaussian identities, the integral in the long-term \TR{expression} can be further simplified as:
\begin{multline}\label{34}
\int\mathcal{N}(\bm{Y}|\bm{K}_{NM}\bm{K}_{M}^{-1}\bm{u}_\mathrm{l}+\bm{m}_\mathrm{s},\sigma_{\epsilon}^2\bm{I}_{N})P\left(\bm{u}_\mathrm{l}\right)\mathrm{d}\bm{u}_\mathrm{l}=\\
\mathcal{N}(\bm{Y}|\bm{m}_\mathrm{s},\bm{Q}_{N}+\sigma_{\epsilon}^2\bm{I}_{N}).
\end{multline}
Furthermore, the KL-divergence between two Gaussians is analytical, that is:
\begin{multline}\label{35}
\mathcal{KL}(q\left(\bm{u}_\mathrm{s}\right)||P\left(\bm{u}_\mathrm{s}\right))=\frac{1}{2}\log|\bm{V}_M|+\frac{1}{2}\bm{m}_{\mathrm{u},\mathrm{s}}^{\top}\bm{V}_M^{-1}\bm{m}_{\mathrm{u},\mathrm{s}}\\
+\frac{1}{2}\mathrm{tr}(\bm{S}_{\mathrm{u},\mathrm{s}}\bm{V}_M^{-1})-\frac{1}{2}M-\frac{1}{2}\log|\bm{S}_{\mathrm{u},\mathrm{s}}|
\end{multline}
Consequently, one gets the analytical form of $\mathcal{L}(q)$:
{\setlength\abovedisplayskip{3pt}
 \setlength\belowdisplayskip{5pt}
\begin{multline}
\label{36}
 \mathcal{L}(q)=\log\mathcal{N}(\bm{Y}|\bm{m}_\mathrm{s},\bm{Q}_{N}+\sigma_{\epsilon}^2\bm{I}_{N})-\frac{1}{2\sigma_{\epsilon}^2}\mathrm{tr}(\bm{K}_{N}-\bm{Q}_{N})\\ 
-\frac{1}{2\sigma_{\epsilon}^2}\mathrm{tr}(\bm{S}_\mathrm{s})-\frac{1}{2}\log|\bm{V}_M|-\frac{1}{2}\bm{m}_{\mathrm{u},\mathrm{s}}^{\top}\bm{V}_M^{-1}\bm{m}_{\mathrm{u},\mathrm{s}}\\ 
-\frac{1}{2}\mathrm{tr}(\bm{S}_{\mathrm{u},\mathrm{s}}\bm{V}_M^{-1})+\frac{1}{2}M+\frac{1}{2}\log|\bm{S}_{\mathrm{u},\mathrm{s}}|.
\end{multline}}
Taking the derivative of $\mathcal{L}(q)$ with respect to the variational parameters of $q(\bm{u}_\mathrm{s})$ and setting them as zeros, we can obtain the optimal $q^{*}(\bm{u}_\mathrm{s})$ with mean and variance:
{\setlength\abovedisplayskip{3pt}
 \setlength\belowdisplayskip{-3pt}
\begin{subequations}
\label{37}
\begin{align}
&\bm{m}_{\mathrm{u},\mathrm{s}}\!=\!\bm{V}_{M}(\bm{V}_{M}\!+\!\bm{V}_{MN}\tilde{\bm{Q}}_{N}^{-1}\bm{V}_{NM})^{-1}\bm{V}_{MN}\tilde{\bm{Q}}_{N}^{-1}\bm{Y}\\[1mm]
&\bm{S}_{\mathrm{u},\mathrm{s}}= \bm{V}_{M}\left(\bm{V}_{M}+\sigma_{\epsilon}^{-2} \bm{V}_{M N} \bm{V}_{N M}\right)^{-1} \bm{V}_{M}
\end{align}
\end{subequations}
}
% \vspace{-1pt}}

\noindent where $\tilde{\bm{Q}}_{N}=\bm{Q}_{N}+\sigma_{\epsilon}^2\bm{I}_{N}$. {Combining~\eqref{28} with the learned variational parameters, one can derive the predictive distribution of both of the GPs:}
{\setlength\abovedisplayskip{3pt}
 \setlength\belowdisplayskip{-1pt}
\begin{subequations}
\label{38}
\begin{align}
q(\YLnew{{\Delta}}_\mathrm{l}^{*}) &= \int P(\YLnew{{\Delta}}_\mathrm{l}^{*}|\bm{u}_\mathrm{l})q^{*}(\bm{u}_\mathrm{l})\mathrm{d}\bm{u}_\mathrm{l}^{*}\triangleq\mathcal{N}(\YLnew{{\mu}}_{\mathrm{l}}^{*},\YLnew{{\sigma}_{\mathrm{l}}^{2*}})\\[-1mm]
q(\YLnew{{\Delta}}_\mathrm{s}^{*}) &= \int P(\YLnew{{\Delta}}_\mathrm{s}^{*}|\bm{u}_\mathrm{s})q^{*}(\bm{u}_\mathrm{s})\mathrm{d}\bm{u}_\mathrm{s}^{*}\triangleq\mathcal{N}(\YLnew{{\mu}}_{\mathrm{s}}^{*},\YLnew{{\sigma}_{\mathrm{s}}^{2*}})
\end{align}
\end{subequations}}

\noindent with $\YLnew{{\mu}}_{\mathrm{l}}^{*}=\bm{K}_{*M}\bm{K}_{M}^{-1}\bm{m}_{\mathrm{u},\mathrm{l}}$, $\YLnew{{\mu}}_{\mathrm{s}}^{*}=\bm{V}_{*M}\bm{V}_{M}^{-1}\bm{m}_{\mathrm{u},\mathrm{s}}$, $\YLnew{{\sigma}_{\mathrm{l}}^{2*}}=k_{**}-\bm{K}_{*M}\bm{K}_{M}^{-1}\bm{K}_{M*}+\bm{K}_{*M}\bm{K}_{M}^{-1}\bm{S}_{\mathrm{u},\mathrm{l}}\bm{K}_{M}^{-1}\bm{K}_{M*}$, and $\YLnew{{\sigma}_{\mathrm{s}}^{2*}}=v_{**}-\bm{V}_{*M}\bm{V}_{M}^{-1}\bm{V}_{M*}+\bm{V}_{*M}\bm{V}_{M}^{-1}\bm{S}_{\mathrm{u},\mathrm{s}}\bm{V}_{M}^{-1}\bm{V}_{M*}$.

Finally, the predictive distribution of the latent function \YLnew{${\Delta}$} at a new test point $\bm{z}^{*}$ is given by
\be
\label{39}
q(\YLnew{{\Delta}}^{*}) = \int P(\YLnew{{\Delta}}^{*}|\YLnew{{\Delta}}_\mathrm{l}^{*},\YLnew{{\Delta}}_\mathrm{s}^{*})q(\YLnew{{\Delta}}_\mathrm{l}^{*})q(\YLnew{{\Delta}}_\mathrm{s}^{*})\mathrm{d}\YLnew{{\Delta}}_\mathrm{l}^{*}\mathrm{d}\YLnew{{\Delta}}_\mathrm{s}^{*}
\ee
which leads to the following predictive mean and variance of the DGP:
\begin{align}
\label{40}
&\YLnew{{\mu}}_{\Delta}(\bm{z}^{*})=\bm{K}_{*M}\bm{K}_{M}^{-1}\bm{m}_{\mathrm{u},\mathrm{l}}+\bm{V}_{*M}\bm{V}_{M}^{-1}\bm{m}_{\mathrm{u},\mathrm{s}}\\ \notag
&\YLnew{{\sigma}^2_{\Delta}}(\bm{z}^{*})=k_{**}+v_{**}-\bm{K}_{*M}\bm{K}_{M}^{-1}\bm{K}_{M*}-\bm{V}_{*M}\bm{V}_{M}^{-1}\bm{V}_{M*}\\ \notag
&\quad+\bm{K}_{*M}\bm{K}_{M}^{-1}\bm{S}_{\mathrm{u},\mathrm{l}}\bm{K}_{M}^{-1}\bm{K}_{M*}+\bm{V}_{*M}\bm{V}_{M}^{-1}\bm{S}_{\mathrm{u},\mathrm{s}}\bm{V}_{M}^{-1}\bm{V}_{M*}
\end{align}
\TR{Note that obtaining these analytic formulas are one of the main contributions of the paper, as the} predictive distribution $q(\YLnew{{\Delta}}^{*})$ of the latent function $\YLnew{{\Delta}}$ can be used for multi-step prediction over a given horizon as it is done in the MPC scheme in the next section. Furthermore, to update the predictive mean and variance of the short-term GP, one can first recursively update the posterior distribution $q(\bm{u}_\mathrm{s})$  with~\eqref{24} using online data points, then marginalize the posterior on $\bm{u}_\mathrm{s}$ as in~\eqref{28}, leading to $q(\YLnew{{\Delta}}^{*})$.

\section{DGP-MPC {Scheme}} \label{sec:4}
\YL{
In this section, we reformulate the MPC problem \eqref{3} by embedding \TR{into the prediction} the proposed dual GP structure \TR{based model, i.e., the predictive posterior distribution}. %into the uncertainty prediction. 
Due to the stochastic property of the dual Gaussian process, \TR{there is an} uncertainty  $\bm{\Delta}(\cdot)$ \TR{associated with the prediction} which \TR{can be} inferred by the DGP structure. % turns into random variables. 
This directly leads to the random variable $\bm{x}_{i|k}$ within the prediction model. Thus, the GP input $\bm{z}$ becomes a random variable during the state propagation, and the posterior of $\bm{\Delta}$. \TR{Under the assumption that}
\begin{multline} \label{41}
{\bm{z}}_{i \mid k}=(\bm{x}_{i \mid k}^{\top}, \bm{u}_{i \mid k}^{\top}) \sim \mathcal{N}\left(\bm{\mu}_i^z, \bm{\Sigma}_i^z\right)\\
=\mathcal{N}\left(\left[\begin{array}{c}
\bm{\mu}_i^x \\
\bm{u}_{i \mid k}
\end{array}\right],\left[\begin{array}{cc}
\bm{\Sigma}_i^x &\mathbf{0} \\
\mathbf{0} & \mathbf{0}
\end{array}\right]\right),
\end{multline}
\TR{the contribution of the GP model to the state-propagation is}
\be
\label{42}
P(\bm{\Delta}_{i|k} | \bm{\mu}_i^z, \bm{\Sigma}_i^z)=\int P(\bm{\Delta}_{i|k} | {\bm{z}}_{i|k}) P({\bm{z}}_{i|k} | \bm{\mu}_i^z, \bm{\Sigma}_i^z) \mathrm{d}{\bm{z}}_{i|k}
\ee
One can observe that the posterior $P(\bm{\Delta}_{i|k} | \bm{\mu}_i^z, \bm{\Sigma}_i^z)$ is a non-Gaussian distribution and intractable to be computed analytically. In this paper, we further extend the exact moment matching approach\cite{quinonero2003prediction,deisenroth2010efficient} into the proposed DGP structure to fit the posterior optimally with the first and second moments of the posterior distribution, that is: $P(\bm{\Delta}_{i|k} | \bm{\mu}_i^z, \bm{\Sigma}_i^z)\approx\mathcal{N}(\bm{\mu}_{\Delta}^{i},\bm{\Sigma}_{\Delta}^{i})$. See Appendix B to find the details about the derivation of \TR{this} Gaussian approximation.}

Then the prediction model in \eqref{3} can be rewritten as $\bm{x}_{i+1 \mid k}\sim\mathcal{N}(\bm{\mu}_x^{i+1}, \bm{\Sigma}_x^{i+1})$ with mean and variance: 
% {\setlength\abovedisplayskip{3pt}
%  \setlength\belowdisplayskip{2pt}
\begin{subequations}
\label{43}
\begin{align}
&\bm{\mu}_x^{i+1}=\bm{A} \bm{\mu}_x^i+\bm{B} \bm{u}_{i \mid k}+\bm{\mu}_{\Delta}^i, \\
&\bm{\Sigma}_x^{i+1}=\bm{A} \bm{\Sigma}_x^i \bm{A}^{\top}+ \bm{\Sigma}_{\Delta}^i + \TR{\bm{\Sigma}_\varepsilon},
\end{align}
\end{subequations}
where $\bm{\Sigma}_{\epsilon}=\diag(\sigma_{\epsilon,1}^2,\ldots,\sigma_{\epsilon,n_\mathrm{x}}^2)$. \YL{On the other hand, the stochastic state also results in a probabilistic cost function $J$. Generally one chooses the expected value to represent the stochastic cost, that is: $\mathbb{E}[J(\bm{x}(k),\bm{U}_k)]$. Recalling the prediction model \eqref{43}, the stochastic cost function can be expressed as a deterministic form:
% {\setlength\abovedisplayskip{3pt}
%  \setlength\belowdisplayskip{3pt}
\begin{multline}\label{44}
\mathbb{E}[J(\bm{x}(k),\bm{U}_k)] 
=\sum_{i=0}^{H-1}\left[l(\bm{\mu}_x^{i},\bm{u}_{i|k})+\operatorname{tr}(\bm{Q} \bm{\Sigma}_x^{i})\right] + \\ l_\mathrm{f}(\bm{\mu}_x^{H})+\operatorname{tr}(\bm{Q}_\mathrm{f} \bm{\Sigma}_x^{H})
\end{multline}}
\YL{
Next, we discuss the satisfaction of the constraints on the state and the control input. As $\bm{x}_{i|k}$ is a random variable with mean $\bm{\mu}_{x}^{i}$ and variance $\bm{\Sigma}_{x}^{i}$,  we intend to satisfy the polytopic constraints in a probabilistic sense, i.e., by the chance constraint\YLnew{\cite{hewing2019cautious,muntwiler2021data,kohler2022recursively}}:
\be
\label{45}
P(\bm{x}_{i|k}\in\mathcal{X})\triangleq P(\bm{C}_x\bm{\mu}^{i}_{x}\leq\bm{c}_x) \geq\gamma,
\ee
where $\gamma$ is \TR{a chosen} probability level of the credibility sets. We utilize the quantile function $\bm{\varpi}(\gamma)$ to convert the above \TR{given} chance constraints into deterministic ones.  Denoting $\bm{e}_j=\bm{C}_{x,j}\bm{x}_{i}-\bm{c}_{x,j}$ with $j=1,...,n_\mathrm{c}$, then we have:
\be
\label{46}
\bm{e}_j\sim\mathcal{N}(\bm{C}_{x,j}\bm{\mu}^{i}_{x}-\bm{c}_{x,j},\bm{C}_{x,j}\bm{\Sigma}_{x}^{i}\bm{C}_{x,j}^{\top}).
\ee
 Thus, the chance constraint \eqref{45} is equivalent to $\bm{\varpi}(\gamma)\leq 0$ and finally leads to:
\be
\label{47}
\bm{C}_{x,j}\bm{\mu}^{i}_{x}\leq \bm{c}_{x,j}-\bm{\varpi}(\gamma)\sqrt{\bm{C}_{x,j}\bm{\Sigma}_{x}^{i}\bm{C}_{x,j}^{\top}}, \quad j\in\TR{\mathbb{I}_1^{n_\mathrm{c}}}.
\ee
Finally, the DGP-MPC problem in this paper is transformed into a deterministic formulation:
\be
{\label{48}
\begin{aligned}
\min_{\bm{U}_k}~&~\eqref{44}\\
{\rm{s.t.}} ~~  
&\eqref{c-2}, \eqref{43}, \eqref{47} \\
&\bm{x}_{0|k}=\bm{x}(k), \bm{u}_{0|k}=\bm{u}(k).
\end{aligned}}
\ee}
Because the stochastic cost has been transformed into a deterministic form, most nonlinear optimization algorithms can be utilized to solve the problem. The specific implementation steps are summarized in Algorithm ~\ref{alg:A}.

\floatname{algorithm}{Algorithm}
\begin{algorithm}
\caption{\YL{Learning based model predictive control strategy with Dual Gaussian process}}
\label{alg:A}
\begin{algorithmic}[1]
% \color{blue}
\Statex \textbf{Initialization}: Constraints $\mathcal{X}$, $\mathcal{U}$, %$\mathcal{D}_\TR{{N}}=\{\}$, \TR{$N=0$}, 
step size~$T_\mathrm{s}$, ~$k=0$,  $\bm{m}_{\mathrm{u},\mathrm{s}}^{0}=\mathbf{0}$, $\bm{S}_{\mathrm{u},\mathrm{s}}^{0}= \beta^{-1}\YLnew{\bm{I}_M}$, $\lambda$, \TR{max.}~number of iterations $\TR{\tau_\mathrm{max}}$.
% \State \textbf{Output}: Control input $\bm{u}$ 
\For{$\TR{\tau}=1,...,\TR{\tau_\mathrm{max}}$}
\Statex \textbf{Offline Phase}:
\State Generate the initial training set  $\mathcal{D}_0$
\State Train the long-term GP to obtain $q\left(\bm{u}_{\mathrm{l}}\right)$ 

\Statex \textbf{Online Phase}:
\For{each time instant $k=1,2,...,$}
     \State Obtain current state $\bm{x}(k)$;
     \State Compute measurement output \YLnew{$\bm{y}(k)$} by \eqref{m-1};
     \State Update $\bm{m}_{\mathrm{u},\mathrm{s}}^{k}$ and  $\bm{S}_{\mathrm{u},\mathrm{s}}^{k}$ using \eqref{22};
        \For{$i=0,...,H$}
        \State State propagation 
        \State $\bm{x}_{i+1|k}\sim\mathcal{N}(\bm{\mu}_x^{i+1},\bm{\Sigma}_x^{i+1})$  
        \State  with \eqref{b-4},~\eqref{b-13}
        \EndFor
     \State Solve \eqref{48} for $\bm{U}_k$;
     \State Apply $\bm{u}(k)\gets \bm{u}_{0|k}$ on the system \eqref{2};
     \State Record data $\mathcal{D}_\TR{k} \gets \mathcal{D}_\TR{k-1}\cup\{\bm{x}(k),~\bm{u}(k)\}$.
\EndFor
\State Set $\mathcal{D}_0\gets\mathcal{D}_\TR{k}$ and go back to offline phase.
\EndFor

%  \State  Add the mission data into $\mathcal{D}_0$ and go back to Step 1. Wait for the next mission.
\end{algorithmic}
\end{algorithm}

\section{Illustrative Examples} \label{sec:5}

\YL{In this section, we present the results and insights gained by applying DGP-MPC Algorithm \ref{alg:A} described in Section 4 for trajectory tracking of \TR{a} quadrotor to illustrate its effectiveness.}

\subsection{Quadrotor dynamics}
The quadrotor is modeled as a rigid body with four rotors. Let $\mathcal{I}$ denote the inertial frame and $\mathcal{B}$ denote the body fixed frame which is attached to the \emph{center of mass} (COM) of the quadrotor and oriented according to $\mathcal{I}$. Define $\bm{p}\in\mathbb{R}^{3}$ and $\bm{v}\in\mathbb{R}^{3}$ as the position and velocity of the COM.
\YL{Denoting $\bm{\zeta}=[\begin{array}{ccc}\phi &\theta &\psi \end{array}]^{\top}$ as the Euler angles of the quadrotor, 
then from the Newton-Euler formulation, one can obtain \TR{a motion} model for the quadrotor:
\be
\label{52}
\begin{aligned}
&\dot{\bm{p}}=\bm{v},  \quad m\dot{\bm{v}}=mg\bm{e}_{3}-T\bm{R}\bm{e}_{3}+\bm{F}_{\Delta},\\
&\dot{\bm{\zeta}}=\bm{\Theta}\bm{\omega},\quad\bm{J}\dot{\bm{\omega}}=-\bm{\omega}^{\times}\bm{J}\bm{\omega}+\bm{\tau}+\bm{\tau}_{\Delta}
\end{aligned}
\ee
where $\bm{R}$ is the rotation matrix from $\mathcal{B}$ to $\mathcal{I}$ with Z-Y-X sequence, $\bm{\omega}\in\mathbb{R}^3$ denotes the angular velocity of $\mathcal{B}$ with respect to $\mathcal{I}$, and %the matrix $\bm{\Theta}\in\mathbb{R}^{3\times 3}$ is described as 
\[\bm{\Theta}= {\tiny\left[\begin{array}{ccc}
c \theta &  0  & -s \theta c \varphi \\
0 &  1 &  s \varphi \\ 
s \theta &  0 & c \theta c \varphi \end{array} \right]}^{-1},\] \TR{with} $s$ and $c$ \TR{being} short hands for sine and cosine, respectively.}
$m=1.9kg$ is the total mass of the quadrotor, 
$\bm{e}_3=[0~0~1]^{\top}$ denotes the unit vector aligning with the gravity $g$ in $\mathcal{I}$, $(\cdot)^{\times}:\mathbb{R}^{3}\to\mathbb{R}^{3\times 3}$ denotes the cross-product operator, $\bm{J}=\mathrm{diag}(5.9,5.9,10.7)\cdot 10^{-3}\mathrm{kg\cdot m^2}$ is the inertia matrix, $T\in\mathbb{R}^{+}$ is the thrust force generated by the four rotors, and $\bm{\tau}\in\mathbb{R}^{3}$ represents the control torque expressed in $\mathcal{B}$. The terms $\bm{F}_{\Delta}$ and $\bm{\tau}_{\Delta}$ represent unknown force and torque \TR{affecting} the quadrotor due to time-varying uncertainties and unmodeled dynamics, such as external wind, complex interactions of rotor airflows affected by the ground and walls, friction and flapping dynamics of the rotor blades. Further details about the rest of the system parameters and the derivation of the terms $\bm{F}_{\Delta}$ and $\bm{\tau}_{\Delta}$  can be found in \cite{kai2017nonlinear}.

\begin{figure}[h]
\begin{center}
  \includegraphics[width= 3.2in]{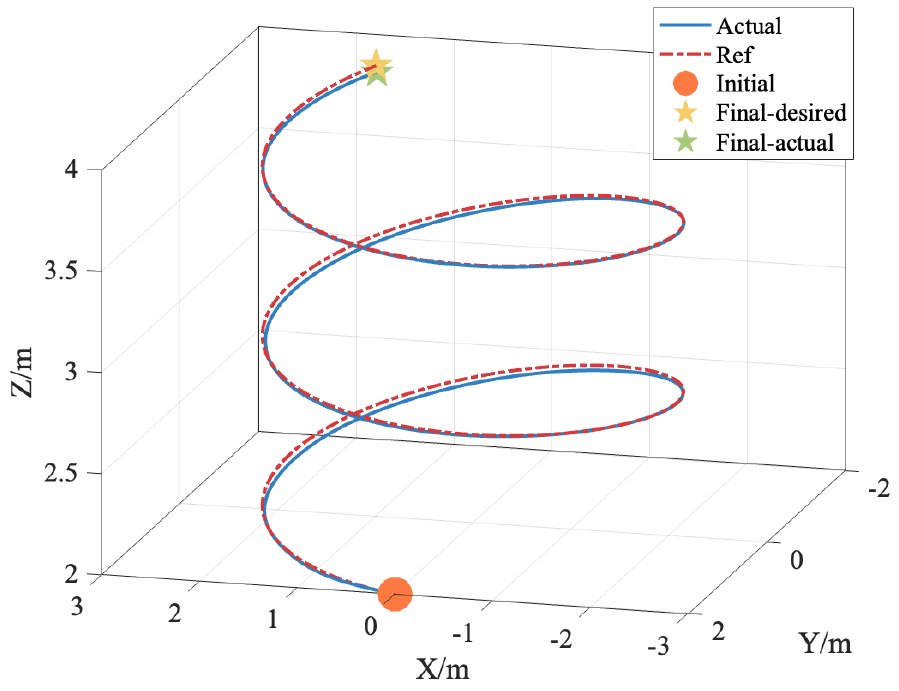}
  \end{center} 
  \caption{\TR{Simulated flight} trajectory with the proposed DGP-MPC \TR{adaptive control method}.} 
  \label{fig:3}
\end{figure} 

By denoting $\bm{x}=[\bm{p}^{\top}\quad\bm{v}^{\top}\quad\bm{\zeta}^{\top}\quad\bm{\omega}^{\top}]^{\top}\in(\mathbb{R}^{12})^{\TR{\mathbb{R}}}$ and $\bm{u}=[T\quad\bm{\tau}^{\top}]^{\top}\in(\mathbb{R}^{4})^{\TR{\mathbb{R}}}$, the continuous-time model~\eqref{52} can be linearized and discretized  resulting in the dynamics given by \eqref{2}.  \YL{The challenges posed by this application are:}

\YL{(1) The \TR{ideal} dynamics \TR{described} by the first principles cannot capture all the unknown internal nonlinearities and external disturbances accurately, especially when there exist time-varying disturbances such as \TR{aerodynamic effects and varying} winds. This is addressed by the recursive online sparse GP method in Section \ref{sec:recGP}.}

\YL{(2) It is not clear how the quadrotor can memorize and exploit historical data when the system experiences a similar environment while performing a repetitive task. This is addressed by the dual GP structure in Section \ref{sec:dualGP}.}

\YL{(3) For the sake of safety, the robust satisfaction of state and input constraints should be guaranteed. Here the state constraints refer to a limited safe space where the quadrotor \TR{can fly} and the input constraints correspond to the maximal thrust force and torque that the quadrotor could provide. This is addressed by the dual GP-MPC strategy in Section \ref{sec:4}.}

\begin{figure}[]
%  \captionsetup{font={small}}
\begin{center}
  \includegraphics[width= 3.5in]{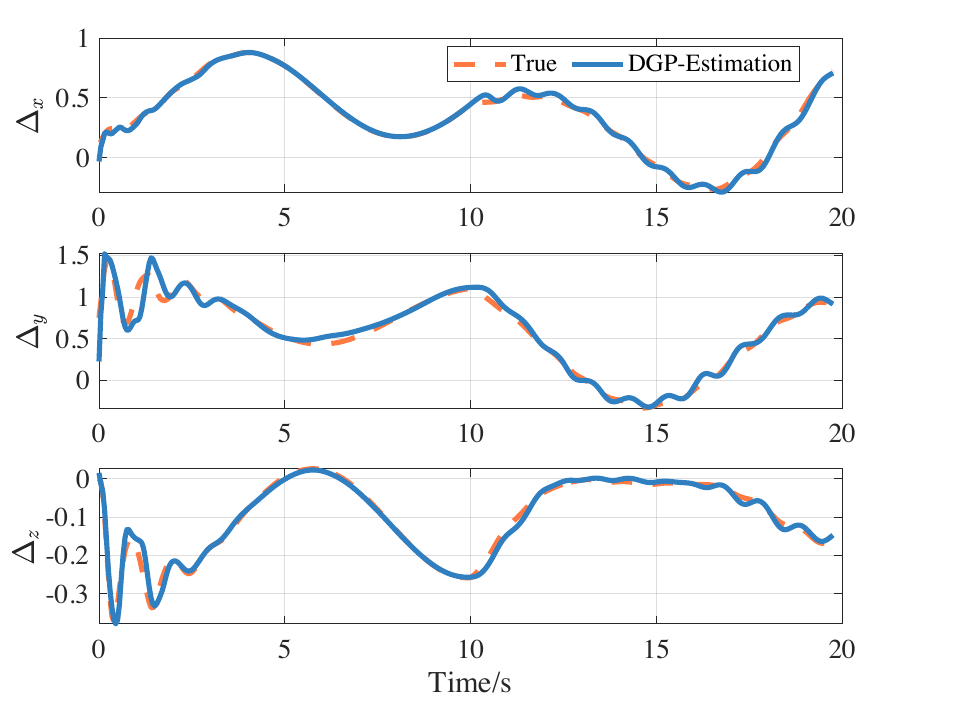}
  \end{center} 
  \caption{\TR{Value of the true uncertain dynamics (given by orange) in terms of $x$,$y$,$z$ components, together with the mean of the learnt DGP structure (given by blue) evaluated at the true state values at time $t$ \YLnew{for the 2nd iteration}.}} 
  \label{fig:4}
\end{figure} 

\begin{figure*}[h]
\begin{center}
  \includegraphics[width= 6.5in]{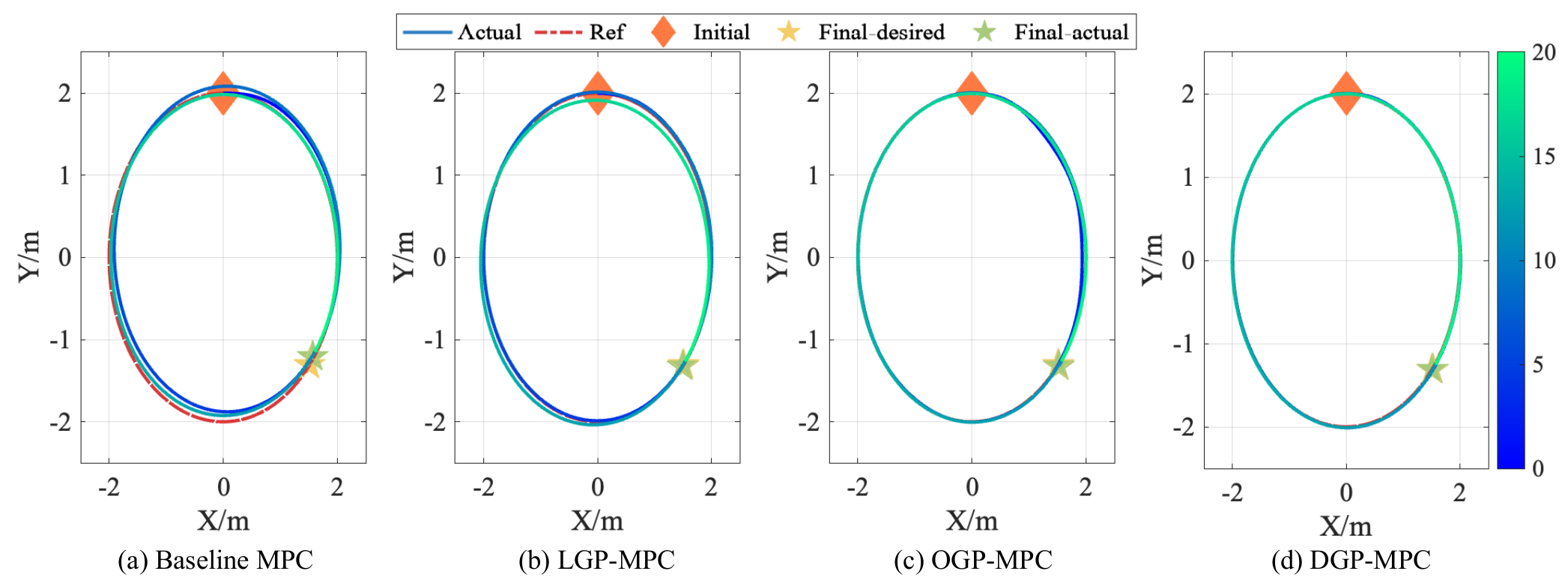}
  \end{center} 
%   \captionsetup{font={small}}
  \caption{\TR{Flight} trajectories \TR{of the quadrotors projected to the} X-Y plane \TR{while being driven by the} controllers. The color of the trajectory indicates the time (from 0-20s).} 
  \label{fig:5}
\end{figure*} 

\subsection{Data Collection}
 \YL{
\TR{Based on our considerations and according} to the disturbance model given in  \cite{kai2017nonlinear}, the GP inputs are designed as $\bm{z}=[\bm{\zeta}^{\top}\quad\bm{v}^{\top}\quad T]^{\top}\in(\mathbb{R}^{7})^\TR{\mathbb{R}}$. Then, to generate the training data set, the quadrotor is enabled to track a random reference trajectory and explore the space. The pseudo-random trajectory is designed as  $x_\mathrm{r}(t)=1.5\sin (t)+1.5\sin(0.33t)$, $y_\mathrm{r}(t)=\sin(1.1t+\pi/2)+\sin(0.11t)$, and $z_\mathrm{r}(t)=\sin(0.33t+\pi/2)+\sin(0.76t)+3$.
It is worth mentioning that the training data can also be collected by flying the quadrotor manually. We use the linear MPC to \TR{approximately follow the reference} by the quadrotor for 50s, and collect the data with 20Hz, resulting in a training data set with size $N = 1000$.  Furthermore, we employ a constant wind with speed $(1~3~-2)^{\top}$ m/s in frame $\mathcal{I}$ which occurs as a heading-dependent uncertainty in \TR{the} dynamics~\eqref{52} during the offline training phase.} \YL{After the data collection phase, the long-term GP is trained with pseudo points \TR{with} size $M=20$,  resulting in the posterior $q(\bm{u}_\mathrm{l})$. Furthermore, the posterior of short-term GP is initialized as $\bm{m}_{\mathrm{u},\mathrm{s}}=\mathbf{0}$, $\bm{S}_{\mathrm{u},\mathrm{s}}=10^{2}\bm{I}$.}

\subsection{Simulation Analysis}
 The quadrotor is forced to track a helix trajectory in 3D space, which is described by: $x_\mathrm{d}\TR{(t)}=2\sin(t)$,  $y_\mathrm{d}\TR{(t)}=2\cos(t)$, and $z_\mathrm{d}\TR{(t)}=2t/T+2$ for \TR{a} total operation time $T=20$s. The sampling time $T_\mathrm{s}$ is 0.05s. Furthermore, to validate the effectiveness of the proposed strategy, the constant wind turns into a time-varying wind at 10s.
 The discrete-time nonlinear system dynamics of the quadrotor are linearized and decomposed into two LTI systems, i.e., the outer-loop translational subsystem and the inner-loop rotational subsystem. To simplify the controller design and speed up the simulation, we adopt a PD controller for the rotational subsystem and use the MPC for the translational subsystem. It is worth mentioning that, despite the linearizing effect \TR{of} feedback control in the rotational subsystem driven by PD controller, the nonlinearities and additional dynamics experienced there affect the translational system, which interprets the necessity of a GP compensation. For the MPC scheme, the prediction horizon is chosen as  $H=5$, and the weight matrices in the cost function are $\bm{Q}=\mathrm{diag}(1, 1, 20, 1, 1, 20)$, and $\bm{R}=\mathrm{diag}(1, 1, 1)$.

The 3D trajectory of \TR{the quadrotor controlled by} the proposed DGP-MPC for the 2nd iteration is intuitively depicted in Fig.~\ref{fig:3}. \YLnew{The mission of trajectory tracking is achieved and }the tracking performance is adequate though the constant wind from 0s - 10s, \TR{which} suddenly turns into the time-varying wind at 10s. \TR{The} learn\TR{ed} unknown function $\bm{\Delta}$ is \TR{depicted} in Fig.~\ref{fig:4}. It is clear that \YLnew{after the 2nd iteration, i.e., with two-task experience, the predictive mean} of the DGP structure \YLnew{evaluated at the true state values at time $t$ }can track the true function in real-time with a fast convergence process because the memory of the trained function is kept in the long-term GP. 

\begin{figure}
%  \captionsetup{font={small}}
\begin{center}
  \includegraphics[width= 3.5in]{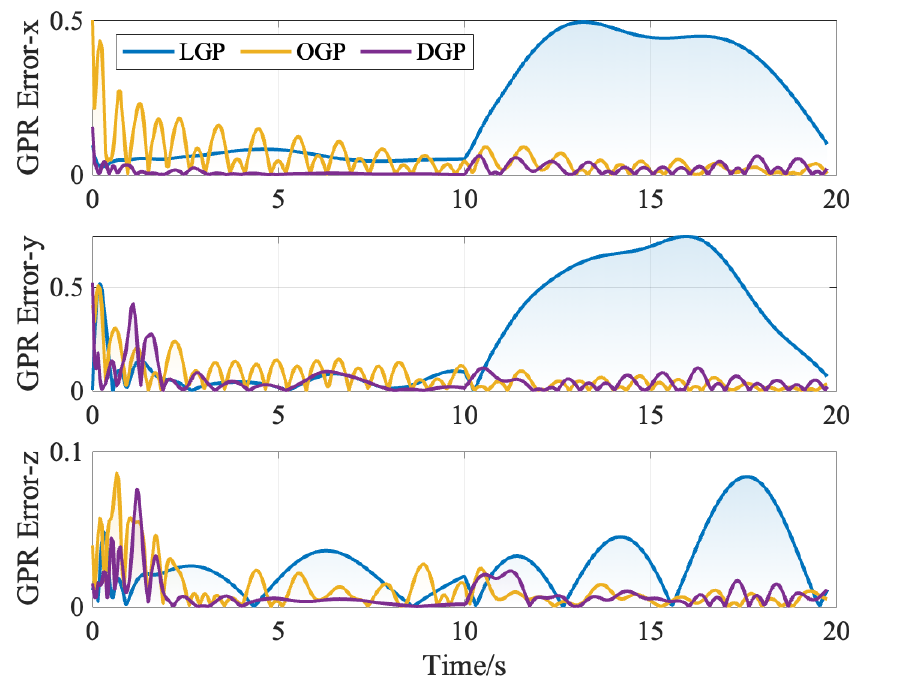}
  \end{center} 
  \caption{Absolute value of the GP estimation error \YLnew{$|{\mu}_{\Delta,i}-\Delta_i|$ for each dimension $i$} \TR{evaluated over time \YLnew{for the 2nd iteration}}.} 
  \label{fig:6}
\end{figure} 

\YL{To show the advantage of the proposed DGP-MPC strategy, the following \TR{methods} are \TR{compared with the DGP solution}: } 

(1) Baseline MPC: MPC strategy \TR{using the baseline linear model} without any GP augmentation;

(2) Long-term only GP-MPC (LGP-MPC): MPC strategy with offline trained GP; 

(3) Online GP-MPC (OGP-MPC): MPC strategy with offline trained GP which \TR{is} recursively update online with \eqref{22}.

To compare the tracking results with 
\TR{these} MPC schemes, Fig.~\ref{fig:5} qualitatively  shows the quadrotor trajectories \TR{by the controllers} \TR{projected to the} X-Y plane, where the color of the trajectory indicates the current time. One can easily notice that, in Fig.~\ref{fig:5}(a) and  Fig.~\ref{fig:5}(b), both the baseline MPC and LGP-MPC have poor dynamic responses during the whole online phase due to their inaccurate prediction model; in Fig.~\ref{fig:5}(c), owing to the recursive online capability of our algorithm, the OGP-MPC shows a relatively better performance than baseline and LGP-MPC even \TR{during the effect of} the unknown time-varying wind. Comparing with Fig.~\ref{fig:5}(c), the \TR{results of the DGP-MPC given} in Fig.~\ref{fig:5}(d) \TR{show} the best control performance during the entire task. The reason is that the proposed dual GP structure enables the real-time accurate compensation of both the system uncertainties and time-varying disturbances. Furthermore, the DGP-MPC preserves the historical data of the repetitive task, which leads to a more precise estimation of the \TR{unknown dynamics} especially in the first half of the task. 

\begin{figure}
%  \captionsetup{font={small}}
\begin{center}
  \includegraphics[width= 3.5in]{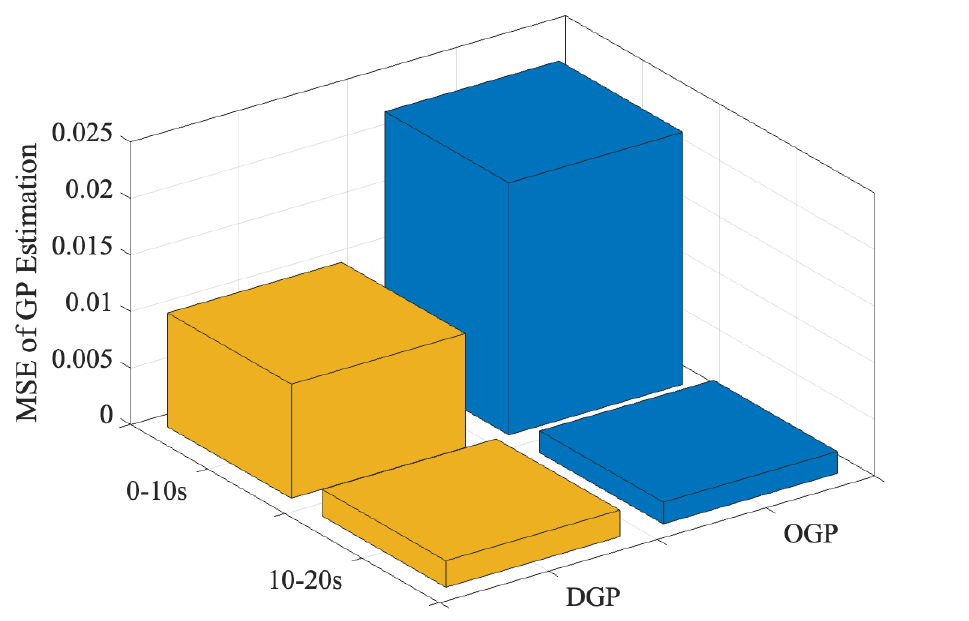}
  \end{center} 
  \caption{MSE of \YLnew{the predictive mean and true function value evaluated at the true state values for both DGP and OGP the 2nd iteration during different phases (constant wind in 0-10s \& time varying wind in 10-20s)}. } 
  \label{fig:7}
\end{figure} 

The ability of \TR{having long-term memory together with active} learning \TR{with} the proposed DGP could also be validated in Figs.~\ref{fig:6}-\ref{fig:7}, which depicts \YLnew{the absolute value of GP estimation error $|{\mu}_{\Delta,i}-\Delta_i|$ }of the LGP, OGP, and DGP in a repetitive task and visualizes the comparison of the \emph{mean squared error} (MSE) \YLnew{of the predictive mean and true function value }for DGP and OGP within the same simulation scenario \TR{through the} different task phases. One can easily deduce that the DGP is capable of estimating the unknown function $\bm{\Delta}$ more accurately within the first 10s of the execution while the LGP results in a dramatic estimation error after $\bm{\Delta}$ begins to vary with time. On the other hand, the OGP needs to re-learn $\bm{\Delta}$ at the beginning since it \TR{inevitably} forgets the previously learned parts of function during the recursive update process. Then, after 10s, \TR{when fast adaptation is required to cope with the time-varying effects} both OGP and DGP \TR{perform equally well. In this case OGP shows a slightly better performance as DGP is more cautious in the adaptation by trying to exploit its long-term memory.}%have the ability to recursive online update, efficient adaptive estimation of $\bm{\Delta}$ can be achieved based on the streaming input and output data.

% In addition, the DGP estimation results and error comparison among the baseline GPMPC, online recursive GPMPC and the proposed DGPMPC are shown in Figs.~\ref{fig:5} - \ref{fig:6}.  It is obvious that both the online GP and DGP can recursively update with the new measurement data after 10s, while the DGP shows better performance ion estimating the uncertain model as it remembers the non-time-varying behavior and focuses the adaptation on the time-varying aspects.

\begin{table}[]
\centering
\caption{Quantitative evaluation of the \TR{performance of the} controllers \TR{in terms of their achieved tracking error.}}
\label{tab:1}
\begin{tabular}{cllll}
\toprule[1.5pt]
\multicolumn{2}{c}{\multirow{2}{*}{Controller}} & \multicolumn{3}{c}{MSE $(\times10^{-
3})$} \\ \cline{3-5} 
\multicolumn{2}{c}{}                            & X      & Y      & Z     \\ \hline
\multicolumn{2}{c}{Baseline MPC}                & 3.30     & 7.71     & 2.93    \\
\multicolumn{2}{c}{LGP-MPC}                      & 0.98     & 2.01     & 0.62    \\
\multicolumn{2}{c}{OGP-MPC}                      & 0.27     & 0.84     & 0.61    \\
\multirow{2}{*}{DGP-MPC}       & 1st iter.       & 0.86     & 0.70     & 0.87    \\
                              & 2nd iter.       & $\textbf{0.07}$     &$\textbf{0.04}$     & $\textbf{0.51}$    \\ \bottomrule[1.5pt]
\end{tabular}
\end{table}

Furthermore, the quantitative evaluation of the proposed DGP-MPC approach in comparison with the baseline controllers is provided in Table.~\ref{tab:1}. One can see that the performance of the proposed DGP-MPC approach \TR{in executing the same trajectory dramatically improves through multiple experiences and after only one attempt it shows much better performance compared to the other control methods}. 

\textbf{Remark 3}. It is worth pointing out that for a non-repetitive task, the proposed DGP-MPC and OGP-MPC have similar control performance due to their abilities of online updat\TR{ing} according to \eqref{22}. However, in terms of the implementation, the posterior $q_k(\bm{\Delta})$ of OGP-MPC with a single GP structure will continuously update with online streaming data, while the initial posterior $q_0(\bm{\Delta})$ obtained from the offline training phase will gradually vanish during the online phase. In contrast, the DGP-MPC proposed in this paper has a dual GP structure with long-term GP and short-term GP. As a part of the overall posterior distribution, the long-term GP posterior with offline trained historical data remains fixed and only the short-term GP posterior is updated with the online data to cope with complex time-varying conditions. Therefore,  the proposed DGP-MPC approach \TR{has a clear advantage} for repetitive tasks.

% One can see that comparing with the baseline controller, the performance of the proposed DGP-MPC strategy in the third iteration has been improved. The reason why the $\bar{J}$ and MSE for baseline GP-MPC are smaller than DGP-MPC in the first iteration is that the estimation of $\bm{\Delta}$ needs to converge to the true value during the first seconds of online recursion in DGP-MPC. Moreover, comparing the $\bar{J}$ and MSE within the iterations of DGP-MPC, one can observe that after incorporating new data into the long-term GP from mission to mission, the tracking error decreases obviously. The reason is that the long-term GP gradually captures more and more model uncertainties after the batch training.

\section{{Conclusions}} \label{sec:6}

\YL{A novel Gaussian Process based \TR{online learning approach for uncertain systems together with an adaptive} model predictive control \TR{scheme have been}  proposed in this paper. Specifically, a dual Gaussian process structure \TR{has been} designed to learn the system uncertainties with both \TR{long-term remembering and rapid} online \TR{learning capabilities}. The knowledge of \TR{reoccurring dynamics are learned and preserved by a} long-term GP while the short-term GP \TR{performs} recursive online adaption to compensate unlearned \TR{or time-varying} uncertainties during the control operation. The features and effectiveness of our approach \TR{have been} illustrated by numerical simulations \TR{and compared}  with multiple MPC schemes. The proposed DGP structure is not only limited to MPC, but also can be extended to any GP-based controller which requires both "memory" and "learning". Future work will consider the generalization of the DGP structure \TR{to} other control approaches}.

\bibliographystyle{unsrt}       
\bibliography{reference}

\appendix
\section{Derivation of variational parameters } \label{app:A}
According to the definition of ELBO, one has:
\begin{align}
{\mathcal{L}}(q)&=\int q(\bm{\Delta},\bm{\Delta}_\mathrm{u})\log \frac{P(\bm{Y},\bm{\Delta},\bm{\Delta}_\mathrm{u})}{q(\bm{\Delta},\bm{\Delta}_\mathrm{u})}\mathrm{d}\bm{\Delta}\mathrm{d}\bm{\Delta}_\mathrm{u}\notag \\
&=\int P(\bm{\Delta}\mid\bm{\Delta}_\mathrm{u})q(\bm{\Delta}_\mathrm{u}) \notag \\ & \hspace{4mm}\log \frac{P(\bm{Y}|\bm{\Delta})\cancel{P(\bm{\Delta}|\bm{\Delta}_\mathrm{u})}P(\bm{\Delta}_\mathrm{u})}{\cancel{P(\bm{\Delta}|\bm{\Delta}_\mathrm{u})}q(\bm{\Delta}_\mathrm{u})}\mathrm{d}\bm{\Delta}\mathrm{d}\bm{\Delta}_\mathrm{u}\notag \\
&=F(q)+\int q(\bm{\Delta}_\mathrm{u})\log \frac{P(\bm{\Delta}_\mathrm{u})}{q(\bm{\Delta}_\mathrm{u})}\mathrm{d}\bm{\Delta}_\mathrm{u} \label{a-1}
\end{align}
where the first term $F(q)$ can be derived as:
\be
\label{a-2}
F(q)=\int q\left(\bm{\Delta}_\mathrm{u}\right)\int P\left(\bm{\Delta}|\bm{\Delta}_\mathrm{u}\right) \log P(\bm{Y} | \bm{\Delta}) \mathrm{d} \bm{\Delta}\mathrm{d} \bm{\Delta}_\mathrm{u}
\ee
The term inside the integral in ~\eqref{a-2} \TR{can be computed as}
\begin{align}
&\int P(\bm{\Delta}\!\mid\!\bm{\Delta}_\mathrm{u})\log P(\bm{Y}|\bm{\Delta})\mathrm{d}\bm{\Delta}= -\frac{|N|}{2}\log 2\pi\!-\!\frac{1}{2}\log |\sigma_{\epsilon}^{2}\bm{I}_{N}| \notag \\
&\quad- \frac{1}{2\sigma_{\epsilon}^{2}}(\bm{Y}\!-\!\bm{K}_{NM}\bm{K}_{M}^{-1}\bm{\Delta}_\mathrm{u})^{\top}(\bm{Y}\!-\!\bm{K}_{NM}\bm{K}_{M}^{-1}\bm{\Delta}_\mathrm{u})\notag \\
&\quad-\frac{1}{2\sigma_{\epsilon}^{2}}\mathrm{tr}\left(\bm{K}_{N}-\bm{Q}_{N}\right)\label{a-3}
\end{align}
where $\bm{Q}_{N}=\bm{K}_{NM}\bm{K}_{M}^{-1}\bm{K}_{MN}$. By benoting $q\left(\bm{\Delta}_\mathrm{u}\right)=\mathcal{N}(\bm{\Delta}_\mathrm{u}|\bm{m}_\mathrm{u},\bm{S}_\mathrm{u})$, we have the analytical form of $F(q)$ as follows:
\be
\label{a-4}
\begin{aligned}
&F(q) =-\frac{|N|}{2}\log 2\pi-\frac{1}{2}\log |\sigma_{\epsilon}^{2}\bm{I}_{N}|-\frac{1}{2\sigma_{\epsilon}^{2}}\bm{Y}^{\top}\bm{Y}\\
&~+\frac{1}{\sigma_{\epsilon}^{2}}\bm{m}_\mathrm{u}^{\top}\bm{K}_{M}^{-1}\bm{K}_{MN}\bm{Y}\!-\!\frac{1}{2\sigma_{\epsilon}^{2}}\bm{m}_\mathrm{u}^{\top}\bm{K}_{M}^{-1}\bm{K}_{MN}\bm{K}_{NM}\bm{K}_{M}^{-1}\bm{m}_\mathrm{u}\\
&~-\frac{1}{2\sigma_{\epsilon}^{2}}\mathrm{tr}(\bm{K}_{M}^{-1}\bm{K}_{MN}\bm{K}_{NM}\bm{K}_{M}^{-1}\bm{S}_\mathrm{u})-\frac{1}{2\sigma_{\epsilon}^{2}}\mathrm{tr}\left(\bm{K}_{N}-\bm{Q}_{N}\right)
\end{aligned}
\ee

Furthermore, the second term in ELBO ~\eqref{a-1} represents the $\mathcal{KL}$ divergence between $q\left(\bm{\Delta}_{{u}}\right)$ and $P\left(\bm{\Delta}_{{u}}\right)$:
\begin{align}
&\int q(\bm{\Delta}_\mathrm{u})\log \frac{P(\bm{\Delta}_\mathrm{u})}{q(\bm{\Delta}_\mathrm{u})}\mathrm{d}\bm{\Delta}_\mathrm{u}\notag\\
&\quad\quad\quad=\mathbb{E}_{q}[\log P(\bm{\Delta}_\mathrm{u})]-\mathbb{E}_{q}[\log q(\bm{\Delta}_\mathrm{u})]\notag\\
&\quad\quad\quad=-\frac{1}{2}\log |\bm{K}_{M}|-\frac{1}{2}\bm{m}_\mathrm{u}^{\top}\bm{K}_{M}^{-1}-\frac{1}{2}\mathrm{tr}\left(\bm{S}_\mathrm{u}\bm{K}_{M}^{-1}\right)\notag\\
&\quad\quad\quad+\frac{|M|}{2}+\frac{1}{2}\log |\bm{S}_\mathrm{u}| \label{a-5}
\end{align}

Based on these derivations, one can readily get the analytical form of the ELBO for the variational distribution $q(\bm{\Delta},\bm{\Delta}_\mathrm{u})$:
\begin{align}
&\mathcal{L}(q)=-\frac{|N|}{2}\log 2\pi-\frac{1}{2}\log |\sigma_{\epsilon}^{2}\bm{I}_{N}|-\frac{1}{2\sigma_{\epsilon}^{2}}\bm{Y}^{\top}\bm{Y}\notag\\
&~+\frac{1}{\sigma_{\epsilon}^{2}}\bm{m}_\mathrm{u}^{\top}\bm{K}_{M}^{-1}\bm{K}_{MN}\bm{Y}\!-\!\frac{1}{2\sigma_{\epsilon}^{2}}\bm{m}_\mathrm{u}^{\top}\bm{K}_{M}^{-1}\bm{K}_{MN}\bm{K}_{NM}\bm{K}_{M}^{-1}\bm{m}_\mathrm{u}\notag\\
&~-\frac{1}{2\sigma_{\epsilon}^{2}}\mathrm{tr}(\bm{K}_{M}^{-1}\bm{K}_{MN}\bm{K}_{NM}\bm{K}_{M}^{-1}\bm{S}_\mathrm{u})-\frac{1}{2\sigma_{\epsilon}^{2}}\mathrm{tr}(\bm{K}_{N}-\bm{Q}_{N})\notag\\
&~-\frac{1}{2}\log |\bm{K}_{M}|-\frac{1}{2}\bm{m}_\mathrm{u}^{\top}\bm{K}_{M}^{-1}-\frac{1}{2}\mathrm{tr}\left(\bm{S}_\mathrm{u}\bm{K}_{M}^{-1}\right)\notag \\
&~+\frac{|M|}{2}+\frac{1}{2}\log |\bm{S}_\mathrm{u}| \label{a-6}
\end{align}

Taking the partial derivatives of ~\eqref{a-6} w.r.t $\bm{m}_\mathrm{u}$ and $\bm{S}_\mathrm{u}$ respectively and set them as zeros, one can finally get the optimal $q^*(\bm{\Delta}_\mathrm{u})$ as \TR{in} \eqref{15}.

\section{Uncertainty propagation for DGP}    % Each appendix must have a short title.
In this section we will derive the mathematical equations necessary for multiple-step ahead prediction with the DGP structure described in Section 3.3.

During the state propagation procedure given in \eqref{43}, the GP input becomes a stochastic variable $P(\bm{z}^*|\tilde{\bm{\mu}},\tilde{\bm{\Sigma}})=\mathcal{N}(\tilde{\bm{\mu}},\tilde{\bm{\Sigma}})$.  This leads to the following predictive distribution:
\be
\label{b-1}
P(\YLnew{{\Delta}}^{*}|\tilde{\bm{\mu}},\tilde{\bm{\Sigma}},\YLnew{\mathcal{D}_N})\!=\!\!\!\int\!\! P(\YLnew{{\Delta}}^{*}|\bm{z}^*,\mathcal{D})P(\bm{z}^*|\tilde{\bm{\mu}},\tilde{\bm{\Sigma}})\mathrm{d}\bm{z}^* 
\ee
Since $P(\YLnew{{\Delta}}^{*}|\bm{z}^*,\YLnew{\mathcal{D}_N})$ is a nonlinear function of $\bm{z}^*$, the predictive distribution \eqref{b-1} is not Gaussian. So it is intractable to obtain the analytical form of the integral \eqref{b-1}. In this paper, we use exact moment matching to make Gaussian approximations to \eqref{b-1}.

First, we rewrite the predictive mean of \TR{the} DGP as:
\be
\label{b-2}
\YLnew{{\mu}}_{\Delta}(\bm{z}^{*})=\bm{K}_{*M}\bm{\beta}_\mathrm{l}+\bm{V}_{*M}\bm{\beta}_\mathrm{s}
\ee
where $\bm{\beta}_\mathrm{l}=\bm{K}_{M}^{-1}\bm{m}_{\mathrm{u},\mathrm{l}}$,  $\bm{\beta}_\mathrm{s}=\bm{V}_{M}^{-1}\bm{m}_{\mathrm{u},\mathrm{s}}$. Then, for predicting at uncertain input   $\bm{z}^*\sim\mathcal{N}(\tilde{\bm{\mu}},\tilde{\bm{\Sigma}})$, we need to compute the first moment mean
\be
\bar{\YLnew{{m}}}(\tilde{\bm{\mu}},\tilde{\bm{\Sigma}})=\mathbb{E}_{\bm{z}^*}[\YLnew{{\mu}}_{\Delta}(\bm{z}^{*})]
\ee
according to the law of iterated expectations. For the convenience of \TR{notation}, we denote the Gaussian kernel \eqref{6} as $\YLnew{\kappa}\left(\bm{z}, \bm{z}'\right)=c\mathcal{N}(\bm{z}|\bm{z}',\bm{\Lambda})$ where $c=\YLnew{\sigma_\mathrm{f}^{2}}(2\pi)^{D/2}|\bm{\Lambda}|^{1/2} $ and $D=n_\TR{\mathrm{x}}+n_\TR{\mathrm{u}}$. Thus, using the product of two Gaussian identities, one has:
\be
\label{b-4}
\begin{aligned}
\bar{\YLnew{{m}}}(\tilde{\bm{\mu}},\tilde{\bm{\Sigma}})
&=\int (\bm{K}_{*M}\bm{\beta}_\mathrm{l}+\bm{V}_{*M}\bm{\beta}_\mathrm{s})P(\bm{z}^*|\tilde{\bm{\mu}},\tilde{\bm{\Sigma}})\mathrm{d}\bm{z}^* \\
&=\sum_{j=1}^{M} {\beta}_{\mathrm{l},j}{L}_{\mathrm{l},j} +\sum_{j=1}^{M} {\beta}_{\mathrm{s},j}{L}_{\mathrm{s},j}\\
&={\bm{\beta}}_{\mathrm{l}}^{\top}{\bm{L}}_{\mathrm{l}}^{(1)}+{\bm{\beta}}_{\mathrm{s}}^{\top}{\bm{L}}_{\mathrm{s}}^{(1)}
\end{aligned}
\ee
where the elements of vector ${\bm{L}}_{\mathrm{l}}^{(1)}$ and ${\bm{L}}_{\mathrm{s}}^{(1)}$ are
\begin{subequations}
\begin{align}
{L}_{\mathrm{l},j}&=c_\mathrm{l}\mathcal{N}(\tilde{\bm{\mu}}|\bm{z}_{\mathrm{u},\mathrm{l}}^{(j)},\bm{\Lambda}_\mathrm{l}+\tilde{\bm{\Sigma}})\\[1mm]
{L}_{\mathrm{s},j}&=c_\mathrm{s}\mathcal{N}(\tilde{\bm{\mu}}|\bm{z}_{\mathrm{u},\mathrm{s}}^{(j)},\bm{\Lambda}_\mathrm{s}+\tilde{\bm{\Sigma}})
\end{align}
\end{subequations}
Next, we compute the variance of the predictive distribution. Similarly, rewriting the variance of \TR{the} DGP in a concise form, one has:
\be
\YLnew{{\sigma}^{2}_{\Delta}}(\bm{z}^{*})=k_{**}+v_{**}-\bm{K}_{*M}\bm{B}_\mathrm{l}\bm{K}_{M*}-\bm{V}_{*M}\bm{B}_\mathrm{s}\bm{V}_{M*}
\ee
where $\bm{B}_\mathrm{l}=\bm{K}_{M}^{-1}-\bm{K}_{M}^{-1}\bm{S}_{\mathrm{u},\mathrm{l}}\bm{K}_{M}^{-1}$ and $\bm{B}_\mathrm{s}=\bm{V}_{M}^{-1}-\bm{V}_{M}^{-1}\bm{S}_{\mathrm{u},\mathrm{s}}\bm{V}_{M}^{-1}$ are the terms independent of test point $\bm{z}^*$.

According to the law of iterated conditional variances, the variance $\bar{\YLnew{{v}}}(\tilde{\bm{\mu}},\tilde{\bm{\Sigma}})$ can be derived as:
\be
\bar{\YLnew{{v}}}(\tilde{\bm{\mu}},\tilde{\bm{\Sigma}}) = \mathbb{E}_{\bm{z}^*}[\YLnew{{\sigma}^{2}_{\Delta}}(\bm{z}^{*})]+\mathbb{E}_{\bm{z}^*}[\YLnew{{\mu}}_{\Delta}(\bm{z}^{*})^2]-\mathbb{E}_{\bm{z}^*}^{2}[\YLnew{{\mu}}_{\Delta}(\bm{z}^{*})]
\ee
where:\\
\begin{align*}
\mathbb{E}_{\bm{z}^*}[\YLnew{{\sigma}^{2}_{\Delta}}(\bm{z}^{*})]&=\sigma_{\mathrm{fl}}^{2}+\sigma_{\mathrm{fs}}^{2} %\notag \\ & \hspace{4mm}
-\sum_{i\in\TR{\mathbb{I}_{1}^N}}\sum_{j\in\TR{\mathbb{I}_{1}^N}}{B}_\mathrm{l}^{i,j}{L}_{\mathrm{l}}^{i,j}-%\sum_{i}\sum_{j}
{B}_\mathrm{s}^{i,j}{L}_{\mathrm{s}}^{i,j}\\
%\end{aligned}
%\ee
%\be
%\begin{aligned}
\mathbb{E}_{\bm{z}^*}[\YLnew{{\mu}}_{\Delta}(\bm{z}^{*})^2]&=\sum_{i\in\TR{\mathbb{I}_{1}^N}}\sum_{j\in\TR{\mathbb{I}_{1}^N}}{\beta}_{\mathrm{l},i}{\beta}_{\mathrm{l},j}{L}_{\mathrm{l}}^{i,j}+%\sum_{i}\sum_{j}
{\beta}_{\mathrm{s},i}{\beta}_{\mathrm{s},j}{L}_{\mathrm{s}}^{i,j}\notag\\
&\hspace{4mm} +2%\sum_{i}\sum_{j}
{\beta}_{\mathrm{l},i}{\beta}_{\mathrm{s},j}\bar{{L}}^{i,j},\\
%\end{aligned}
%\ee
%\be
%\begin{aligned}
\mathbb{E}_{\bm{z}^*}^{2}[\YLnew{{\mu}}_{\Delta}(\bm{z}^{*})]&=\bar{\YLnew{{m}}}(\tilde{\bm{\mu}},\tilde{\bm{\Sigma}})^2.
\end{align*}
Therefore, 
\begin{align}
{L}_{(\cdot)}^{i,j} &= c_{(\cdot)}^2\mathcal{N}(\bm{z}_{\mathrm{u}(\cdot)}^i|\bm{z}_{\mathrm{u}(\cdot)}^j,2\bm{\Lambda}_{(\cdot)})\mathcal{N}(\tilde{\bm{\mu}}|\bm{z}_{d(\cdot)},\tilde{\bm{\Sigma}}+\frac{1}{2}\bm{\Lambda}_{(\cdot)}), \\
\bar{{L}}^{i,j} &= c_\mathrm{l}^2c_\mathrm{s}^2\mathcal{N}(\bm{z}_{\mathrm{u},\mathrm{l}}^i|\bm{z}_{\mathrm{u},\mathrm{s}}^j,\bm{\Lambda}_{\mathrm{l}}+\bm{\Lambda}_{\mathrm{s}})\mathcal{N}(\tilde{\bm{\mu}}|\bm{\xi}_{i,j},\bm{\Xi}).
\end{align}
where $\bm{z}_{d(\cdot)}=(\bm{z}_{\mathrm{u}(\cdot)}^i+\bm{z}_{\mathrm{u}(\cdot)}^j)/2$, $\bm{\xi}_{i,j}=\bm{\Lambda}_{\mathrm{s}}(\bm{\Lambda}_{\mathrm{l}}+\bm{\Lambda}_{\mathrm{s}})^{-1}\bm{z}_{\mathrm{u},\mathrm{l}}^i+\bm{\Lambda}_{\mathrm{l}}(\bm{\Lambda}_{\mathrm{l}}+\bm{\Lambda}_{\mathrm{s}})^{-1}\bm{z}_{\mathrm{u},\mathrm{s}}^j$, $\bm{\Xi}=(\bm{\Lambda}_{\mathrm{l}}^{-1}+\bm{\Lambda}_{\mathrm{s}}^{-1})^{-1}+\tilde{\bm{\Sigma}}$. This leads to the  solution for the predictive variance of:
\begin{align}
\bar{\YLnew{{v}}}(\tilde{\bm{\mu}},\tilde{\bm{\Sigma}}) &= \sigma_{\mathrm{fl}}^{2}+\sigma_{\mathrm{fs}}^{2}-\mathrm{tr}((\bm{B}_\mathrm{l}-\bm{\beta}_\mathrm{l}\bm{\beta}_\mathrm{l}^{\top})\bm{L}_{\mathrm{l}}^{(2)})\notag \\
&~~-\mathrm{tr}((\bm{B}_\mathrm{s}-\bm{\beta}_\mathrm{s}\bm{\beta}_\mathrm{s}^{\top})\bm{L}_{\mathrm{s}}^{(2)})+2\mathrm{tr}(\bm{\beta}_\mathrm{l}\bm{\beta}_\mathrm{s}^{\top}\bar{\bm{L}})\notag \\
&~~-\bar{\YLnew{{m}}}(\tilde{\bm{\mu}},\tilde{\bm{\Sigma}})^2. \label{b-13}
\end{align}

It is worth mentioning that, if the GP input is deterministic without any uncertainty, i.e., $\tilde{\bm{\Sigma}}=\mathbf{0}$, then $\bar{\YLnew{{m}}}(\tilde{\bm{\mu}},\mathbf{0})$ tends to $\YLnew{{\mu}}_{\Delta}(\bm{z}^{*})$ and $\bar{\YLnew{{v}}}(\tilde{\bm{\mu}},\mathbf{0})$ collapses to $\YLnew{{\sigma}^{2}_{\Delta}}(\bm{z}^{*})$ as we expect.

\end{document}